\documentclass[twocolumn,twocolappendix]{aastex63}

\graphicspath{{./}{figures/}}
\usepackage{color}
\usepackage{soul}
\usepackage{natbib}
\usepackage{hyperref}
\usepackage{amsmath}
\usepackage{xspace}
\usepackage{xcolor}
\usepackage[dvipsnames]{xcolor}
\usepackage{graphicx}
\usepackage{enumitem}
\usepackage{amssymb}
\usepackage{xifthen}
\usepackage{hyperref}
\usepackage[normalem]{ulem}
\usepackage{comment}
\usepackage{float}
\usepackage{CJKutf8}
\usepackage{bbm}
\hypersetup{    
  colorlinks      = {true},
  linkcolor       = {blue},
  citecolor       = {blue},
  urlcolor        = {blue},
}

\newcommand{\code}[1]{\texttt{#1}\xspace}

\newcommand{\teff}{\ensuremath{T_\mathrm{eff}}\xspace}

\usepackage{graphicx,pifont}% http://ctan.org/pkg/{graphicx,pifont}
\let\oldding\ding% Store old \ding in \oldding
\renewcommand{\ding}[2][1]{\scalebox{#1}{\oldding{#2}}}% Scale \oldding via optional argument

\usepackage{array}
\newcolumntype{H}{>{\setbox0=\hbox\bgroup}c<{\egroup}@{}}

\shorttitle{The metallicity of DESI Y1 RR Lyrae}
\shortauthors{Medina, Li, Allende-Prieto, et al.}
\begin{document}

\title{The DESI Y1 RR Lyrae catalog II:\\ 
The metallicity dependency of pulsational properties, \\the shape of the RR Lyrae instability strip, and metal rich RR Lyrae} 

\author[0000-0003-0105-9576]{Gustavo~E.~Medina}
\affiliation{David A. Dunlap Department of Astronomy \& Astrophysics, University of Toronto, 50 St George Street, Toronto ON M5S 3H4, Canada}
\affiliation{Dunlap Institute for Astronomy \& Astrophysics, University of Toronto, 50 St George Street, Toronto, ON M5S 3H4, Canada}

\author[0000-0002-9110-6163]{Ting~S.~Li}
\affiliation{David A. Dunlap Department of Astronomy \& Astrophysics, University of Toronto, 50 St George Street, Toronto ON M5S 3H4, Canada}
\affiliation{Dunlap Institute for Astronomy \& Astrophysics, University of Toronto, 50 St George Street, Toronto, ON M5S 3H4, Canada}

% alphabetical:

\author[0000-0002-0084-572X]{C.~Allende~Prieto}
\affiliation{Departamento de Astrof\'{\i}sica, Universidad de La Laguna (ULL), E-38206, La Laguna, Tenerife, Spain}
\affiliation{Instituto de Astrof\'{\i}sica de Canarias, C/ V\'{\i}a L\'{a}ctea, s/n, E-38205 La Laguna, Tenerife, Spain}

\author[0000-0002-0740-1507]{L. {Beraldo e Silva}}
\affiliation{Steward Observatory, University of Arizona, 933 N, Cherry Ave, Tucson, AZ 85721, USA}
\affiliation{Observatório Nacional, Rio de Janeiro - RJ, 20921-400, Brasil}
%\author[0000-0002-0740-1507]{L. Beraldo e Silva}
%\affiliation{Department of Astronomy, University of Michigan, Ann Arbor, MI 48109, USA}

\author[0000-0002-5689-8791]{A.~Bystr\"om}
\affiliation{Institute for Astronomy, University of Edinburgh, Royal Observatory, Blackford Hill, Edinburgh EH9 3HJ, UK}

\author[0000-0002-7667-0081]{R.~G.~Carlberg}
\affiliation{Department of Astronomy \& Astrophysics, University of Toronto, Toronto, ON M5S 3H4, Canada}

\author[0000-0003-2644-135X]{S.~E.~Koposov}
\affiliation{Institute for Astronomy, University of Edinburgh, Royal Observatory, Blackford Hill, Edinburgh EH9 3HJ, UK}
\affiliation{Institute of Astronomy, University of Cambridge, Madingley Road, Cambridge CB3 0HA, UK}

\author[0000-0002-2527-8899]{M. Lambert}
\affiliation{Department of Astronomy \& Astrophysics, University of California, Santa Cruz, Santa Cruz, CA 95064, USA}

%\author[0000-0001-5805-5766]{A.~H.~Riley}
%\affiliation{Institute for Computational Cosmology, Department of Physics, Durham University, South Road, Durham DH1 3LE, UK}

\author[0000-0002-5758-150X]{J.~R.~Najita}
\affiliation{NSF NOIRLab, 950 N. Cherry Avenue, Tucson, AZ 85719, USA}

\author[0000-0002-6667-7028]{C.~Rockosi}
\affiliation{Department of Astronomy and Astrophysics, UCO/Lick Observatory, University of California, 1156 High Street, Santa Cruz, CA 95064, USA}
\affiliation{Department of Astronomy and Astrophysics, University of California, Santa Cruz, 1156 High Street, Santa Cruz, CA 95065, USA}
\affiliation{University of California Observatories, 1156 High Street, Sana Cruz, CA 95065, USA}

\author{N.~Kizhuprakkat}
\affiliation{Institute of Astronomy and Department of Physics, National Tsing Hua University, 101 Kuang-Fu Rd. Sec. 2, Hsinchu 30013, Taiwan}

\author[0000-0001-5805-5766]{A.~H.~Riley}
\affiliation{Institute for Computational Cosmology, Department of Physics, Durham University, South Road, Durham DH1 3LE, UK}

%\author[0009-0001-7217-8006]{J. Yu}
%\affiliation{Institute of Physics, Laboratory of Astrophysics, \'Ecole Polytechnique F\'ed\'erale de Lausanne (EPFL), Observatoire de Sauverny, Chemin Pegasi 51, CH-1290 Versoix, Switzerland}

%\author[0000-0003-0853-8887]{N.~Kizhuprakkat}
%\affiliation{Institute of Astronomy and Department of Physics, National Tsing Hua University, Hsinchu 30013, Taiwan}
%\affiliation{Center for Informatics and Computation in Astronomy, National Tsing Hua University, Hsinchu 30013, Taiwan}

%\author[0000-0002-6257-2341]{M.~Valluri}
%\affiliation{Department of Astronomy, University of Michigan, Ann Arbor, MI 48109, USA}
%\affiliation{University of Michigan, Ann Arbor, MI 48109, USA}

%\author[0000-0002-5762-7571]{W.~Wang}
%\affiliation{Department of Astronomy, School of Physics and Astronomy, and Shanghai Key Laboratory for Particle Physics and Cosmology, Shanghai Jiao Tong University, Shanghai 200240, People’s Republic of China}
%\affiliation{State Key Laboratory of Dark Matter Physics, School of Physics and Astronomy,Shanghai Jiao Tong University, Shanghai 200240, China}
%\affiliation{Shanghai Key Laboratory for Particle Physics and Cosmology, Shanghai 200240, China}

%\author[0000-0001-9852-9954]{O.~Y.~Gnedin}
%\affiliation{University of Michigan, Ann Arbor, MI 48109, USA}

% Author list file generated with: mkauthlist 1.3.0 
% mkauthlist -f --orcid DESI-2025-0522_author_list_2ndtier.csv example_author_list_2ndtier.tex 

\author{J.~Aguilar}
\affiliation{Lawrence Berkeley National Laboratory, 1 Cyclotron Road, Berkeley, CA 94720, USA}
\author[0000-0001-6098-7247]{S.~Ahlen}
\affiliation{Department of Physics, Boston University, 590 Commonwealth Avenue, Boston, MA 02215 USA}
\author[0000-0001-9712-0006]{D.~Bianchi}
\affiliation{Dipartimento di Fisica ``Aldo Pontremoli'', Universit\`a degli Studi di Milano, Via Celoria 16, I-20133 Milano, Italy}
\affiliation{INAF-Osservatorio Astronomico di Brera, Via Brera 28, 20122 Milano, Italy}
\author{D.~Brooks}
\affiliation{Department of Physics \& Astronomy, University College London, Gower Street, London, WC1E 6BT, UK}
\author{T.~Claybaugh}
\affiliation{Lawrence Berkeley National Laboratory, 1 Cyclotron Road, Berkeley, CA 94720, USA}
\author[0000-0001-8274-158X]{A.~P.~Cooper}
\affiliation{Institute of Astronomy and Department of Physics, National Tsing Hua University, 101 Kuang-Fu Rd. Sec. 2, Hsinchu 30013, Taiwan}
\author[0000-0002-1769-1640]{A.~de la Macorra}
\affiliation{Instituto de F\'{\i}sica, Universidad Nacional Aut\'{o}noma de M\'{e}xico,  Circuito de la Investigaci\'{o}n Cient\'{\i}fica, Ciudad Universitaria, Cd. de M\'{e}xico  C.~P.~04510,  M\'{e}xico}
\author[0000-0002-4928-4003]{A.~Dey}
\affiliation{NSF NOIRLab, 950 N. Cherry Ave., Tucson, AZ 85719, USA}
\author{P.~Doel}
\affiliation{Department of Physics \& Astronomy, University College London, Gower Street, London, WC1E 6BT, UK}
\author[0000-0002-2890-3725]{J.~E.~Forero-Romero}
\affiliation{Departamento de F\'isica, Universidad de los Andes, Cra. 1 No. 18A-10, Edificio Ip, CP 111711, Bogot\'a, Colombia}
\affiliation{Observatorio Astron\'omico, Universidad de los Andes, Cra. 1 No. 18A-10, Edificio H, CP 111711 Bogot\'a, Colombia}
\author{E.~Gaztañaga}
\affiliation{Institut d'Estudis Espacials de Catalunya (IEEC), c/ Esteve Terradas 1, Edifici RDIT, Campus PMT-UPC, 08860 Castelldefels, Spain}
\affiliation{Institute of Cosmology and Gravitation, University of Portsmouth, Dennis Sciama Building, Portsmouth, PO1 3FX, UK}
\affiliation{Institute of Space Sciences, ICE-CSIC, Campus UAB, Carrer de Can Magrans s/n, 08913 Bellaterra, Barcelona, Spain}
\author[0000-0003-3142-233X]{S.~Gontcho A Gontcho}
\affiliation{Lawrence Berkeley National Laboratory, 1 Cyclotron Road, Berkeley, CA 94720, USA}
\author{G.~Gutierrez}
\affiliation{Fermi National Accelerator Laboratory, PO Box 500, Batavia, IL 60510, USA}
\author[0000-0002-6024-466X]{M.~Ishak}
\affiliation{Department of Physics, The University of Texas at Dallas, 800 W. Campbell Rd., Richardson, TX 75080, USA}
\author{R.~Kehoe}
\affiliation{Department of Physics, Southern Methodist University, 3215 Daniel Avenue, Dallas, TX 75275, USA}
\author[0000-0003-3510-7134]{T.~Kisner}
\affiliation{Lawrence Berkeley National Laboratory, 1 Cyclotron Road, Berkeley, CA 94720, USA}
\author[0000-0003-1838-8528]{M.~Landriau}
\affiliation{Lawrence Berkeley National Laboratory, 1 Cyclotron Road, Berkeley, CA 94720, USA}
\author[0000-0001-7178-8868]{L.~Le~Guillou}
\affiliation{Sorbonne Universit\'{e}, CNRS/IN2P3, Laboratoire de Physique Nucl\'{e}aire et de Hautes Energies (LPNHE), FR-75005 Paris, France}
\author[0000-0002-1125-7384]{A.~Meisner}
\affiliation{NSF NOIRLab, 950 N. Cherry Ave., Tucson, AZ 85719, USA}
\author{R.~Miquel}
\affiliation{Instituci\'{o} Catalana de Recerca i Estudis Avan\c{c}ats, Passeig de Llu\'{\i}s Companys, 23, 08010 Barcelona, Spain}
\affiliation{Institut de F\'{i}sica d’Altes Energies (IFAE), The Barcelona Institute of Science and Technology, Edifici Cn, Campus UAB, 08193, Bellaterra (Barcelona), Spain}
\author[0000-0001-7145-8674]{F.~Prada}
\affiliation{Instituto de Astrof\'{i}sica de Andaluc\'{i}a (CSIC), Glorieta de la Astronom\'{i}a, s/n, E-18008 Granada, Spain}
\author[0000-0001-6979-0125]{I.~P\'erez-R\`afols}
\affiliation{Departament de F\'isica, EEBE, Universitat Polit\`ecnica de Catalunya, c/Eduard Maristany 10, 08930 Barcelona, Spain}
\author{G.~Rossi}
\affiliation{Department of Physics and Astronomy, Sejong University, 209 Neungdong-ro, Gwangjin-gu, Seoul 05006, Republic of Korea}
\author[0000-0002-9646-8198]{E.~Sanchez}
\affiliation{CIEMAT, Avenida Complutense 40, E-28040 Madrid, Spain}
\author{D.~Schlegel}
\affiliation{Lawrence Berkeley National Laboratory, 1 Cyclotron Road, Berkeley, CA 94720, USA}
\author[0000-0002-3461-0320]{J.~Silber}
\affiliation{Lawrence Berkeley National Laboratory, 1 Cyclotron Road, Berkeley, CA 94720, USA}
\author{D.~Sprayberry}
\affiliation{NSF NOIRLab, 950 N. Cherry Ave., Tucson, AZ 85719, USA}
\author[0000-0003-1704-0781]{G.~Tarl\'{e}}
\affiliation{University of Michigan, 500 S. State Street, Ann Arbor, MI 48109, USA}
\author{B.~A.~Weaver}
\affiliation{NSF NOIRLab, 950 N. Cherry Ave., Tucson, AZ 85719, USA}
\author[0000-0001-5381-4372]{R.~Zhou}
\affiliation{Lawrence Berkeley National Laboratory, 1 Cyclotron Road, Berkeley, CA 94720, USA}

%\author{the DESI collaboration}

\correspondingauthor{Gustavo E. Medina}
\email{gustavo.medina@utoronto.ca}

% Enter the current year, for the copyright statements etc.
%\pubyear{2023}

\begin{abstract}
RR Lyrae stars (RRLs) are valuable probes of both Milky Way assembly and stellar-evolution physics. Using a sample 6,240 RRLs obtained in the first year of the Dark Energy Spectroscopic Instrument (DESI) survey, we investigate the metallicity of RRLs and its correlation with their pulsation properties.
We find that 
(1) a clear correlation between period and [Fe/H] reinforces the view that the long-standing Oosterhoff dichotomy arises from the scarcity of intermediate-metallicity Galactic globular clusters hosting sizeable RRL samples; 
(2) high-amplitude short-period and small-amplitude short-period variables are comparatively metal-rich, with mean [Fe/H] = $-1.39 \pm 0.27$ and $-1.30 \pm 0.28$, respectively; 
(3) in double-mode pulsators (RRd) the metallicity declines smoothly with increasing fundamental-mode period, 
and anomalous RRd stars occupy a remarkably narrow [Fe/H] range relative to classical RRd stars; 
(4) this spectroscopic sample let us, for the first time, place empirical constraints 
on the metallicity-dependent 
topology of the instability strip using phase-corrected effective temperatures and a large number of RRLs, 
where we observe an instability strip that moves towards cooler \teff\ with declining [Fe/H] with a width roughly consistent with stellar-evolution models;
and 
(5) a subset of metal-rich RRLs exhibits orbits consistent with disk membership and  halo kinematics. 
Our results confirm the tantalizing potential of DESI for Galactic and stellar astrophysics and highlight the importance of the even larger samples of RRLs and data-processing improvements forthcoming in future DESI data releases.
\end{abstract}

%\begin{keywords}
%\end{keywords}

% https://astrothesaurus.org/concept-select/

%%%%%%%%%%%%%%%%%%%%%%%%%%%%%%%%%%%%%%%%%%%%%%%%%%
%%%%%%%%%%%%%%%%%%%%%%%%%%%%%%%%%%%%%%%%%%%%%%%%%%
%%%%%%%%%%%%%%%%%%%%%%%%%%%%%%%%%%%%%%%%%%%%%%%%%%

%%%%%%%%%%%%%%%%% BODY OF PAPER %%%%%%%%%%%%%%%%%%

%%%%%%%%%%%%%%%%%%%%%%%%%%%%%%%%%%%%%%%%%%%%%%%%%%
%%%%%%%%%%%%%%%%%%%%%%%%%%%%%%%%%%%%%%%%%%%%%%%%%%
%%%%%%%%%%%%%%%%%%%%%%%%%%%%%%%%%%%%%%%%%%%%%%%%%%

\section{Introduction} \label{sec:intro}

RR Lyrae stars (hereafter RRLs; RRL for a single star) play a pivotal role in unraveling the formation history of our Galaxy, the Milky Way, and serve as key benchmarks for advancing our understanding of stellar evolution \citep[][]{Bono1994,Catelan2009}.  
These low-mass variable stars are predominantly old and metal-poor (Population II) stars, commonly found in the Galactic halo \citep[e.g.,][]{Zinn2014,Torrealba2015,Sesar2017,Medina2018,Stringer2021,Medina2024}, globular clusters \citep[e.g.,][]{Bailey1913,Castellani2003,Navarrete2015,Kundu2019,Cruz-Reyes2024}, and dwarf galaxies \citep[e.g.,][]{Medina2017,Martinez-Vazquez2019,Torrealba2019,Vivas2020,Monelli2022}, and are widely known as precise distance indicators \citep[][]{Catelan2015,Beaton2018}. 
They lie in the intersection between the horizontal branch and instability strip in the Hertzprung-Russel (HR) diagram, and are characterized by their short periods of pulsation ($<$1\,d).  

RR Lyrae variables can be divided into three main subclasses. 
The most numerous among different RRL classes pulsate in the fundamental mode (ab-type RRLs; RRab), with saw-tooth shaped light curves, periods ranging from 0.4\,d to 1.0\,d, and amplitudes that anti-correlate with period and reach up to $\sim1.2$\,mag in the $V-$band. 
RR Lyrae variables that pulsate in the first-overtone mode, (c-type RRLs; RRc),  display shorter periods than RRab stars (0.25-0.40\,d) with smaller amplitudes ($<0.7$\,mag in the $V-$band). 
The third subclass (d-type RRLs; RRd) is significantly less common than RRab and RRc. 
Stars in this class pulsate in both the fundamental and the first-overtone mode, the latter being the dominant one \citep[][]{Nemec1985,Bono1996,Nemec2024}. 
For most RRd stars, the first-overtone pulsation displays larger amplitude than the fundamental mode component, with first-overtone light curves that resemble those of single-mode RRc stars. 
The position of the RRL subtypes in the HR diagram is a good diagnostic tool for constraining the shape of the instability strip \citep[see][and references therein]{Catelan2015}. 
Indeed, RRab stars are located closer to the cooler (red) edge of the instability strip, whereas RRc stars lie towards its hotter (blue) edge and RRd stars lie in the intermediate region \citep[see e.g.,][]{Bono1994}.

The chemical abundance patterns of RRLs, and iron abundance in particular, play an important role in their evolution and their use as stellar population tracers in the local universe. 
For instance, what makes RRLs excellent distance indicators is the tight correlation between their periods of pulsation, metallicities, and luminosity in the infrared, and their luminosity-metallicity relation in optical bands. 
These correlations have been extensively studied empirically \citep[e.g.,][]{Muraveva2015,Sesar2017,Garofalo2022,Prudil2024a,Narloch2024} and with the use of stellar evolution models \citep[e.g.,][]{Bono2003,Catelan2004,Neeley2017,Marconi2022}, and their calibration relies on the choice of a given metallicity scale. 
The iron abundance of RRLs also plays a role in the shape of their light curves, as measured by their Fourier decomposition, which is reflected in known correlations between the phase parameter $\phi_{31}$ (defined using the first and third Fourier coefficients, $\phi_1$ and $\phi_3$), their periods, and [Fe/H] \citep[see e.g.,][]{Jurcsik1996,Smolec2005,Nemec2013,Dekany2021,Mullen2021,Mullen2022}.
Moreover, the morphology of the horizontal branch, and therefore, the depth of its incursion into the instability strip (at a fixed age) is highly dependent on metallicity \citep[see e.g.,][]{Savino2020}, and even the 
topology of the instability strip is a function of metallicity \citep[][]{Marconi2015}. 
Furthermore, the smooth gradient in metallicity as a function of period observed among field RRLs \citep[][]{Fabrizio2019,Fabrizio2021} and the link between (high metallicity) short-period RRLs and massive Milky Way satellites \citep[][]{Fiorentino2015,Monelli2022} provide valuable insights into the origin of the halo. 
Additionally, the presence of RRLs with anomalously high [Fe/H] in the Milky Way might suggest the existence of alternative formation mechanisms for these stars \citep[e.g.,][]{Bono1997a,Bono1997b,Bobrick2024}. 
A major limitation in performing statistically robust empirical studies in any of these research areas is the strong requirement for access to sizable samples of RRLs with uniformly determined metallicities.

The Dark Energy Spectroscopic Instrument \citep[DESI;][]{DESI2016a, DESI_Instrument_Design_2016,DESI_Instrument_2022} Milky Way survey provides a unique dataset to study the spectroscopic properties of RRLs, and, in particular, their chemical abundances. 
DESI is mounted at the Mayall 4-m telescope at Kitt Peak National Observatory and is composed of 5000 robotic fibers that feed ten thermally-controlled 3-channel spectrographs, covering a wavelength range of 3600 to 9800\,\AA\ with a resolution of $\sim2000-5000$ ($\sim1.8$\,\AA).
While DESI's main scientific goal is to perform cosmological studies of the large-scale structure of the universe and its expansion \citep[see e.g.,][]{Levi2013,DESI_DR1_cosmology_2024,DESI_DR2_cosmology_2025}, DESI has also observed several millions of stars in the Galactic disk and the halo as part of the bright-time Milky Way survey \citep[see e.g.,][]{Cooper2023, Koposov2024}. 
Indeed, recent works that reveal the tremendous potential of DESI for Milky Way science include the use of blue horizontal-branch (BHB) stars \citep[][]{Bystrom2024} to characterize the bulk motion of the Milky Way disk caused by the influence of the Large Magellanic Cloud (LMC), the study of nearby substructures in the Galactic halo \citep[][]{Kim2025}, the compilation of white dwarf catalogs and identification of new white dwarf classes \citep[][]{Manser2023,Manser2024}, and the identification of metal-poor stars \citep[][]{Allende-Prieto2023}. 
More specifically, thousands of RRLs with homogeneously-derived spectroscopic properties ([Fe/H], \teff, $\log g$, and center-of-mass radial velocities) were observed in DESI's first year of operation. 
A detailed analysis of the DESI Year 1 (Y1) RRL catalog, which contains thousands of RRab and RRc stars, and hundreds of RRd stars, is presented in a companion paper \citep[][hereafter M25]{Medina2025}

In this paper, we use the [Fe/H] measurements in the DESI Y1 database (M25) to study the metallicity of RRLs and its correlation with their periods and amplitudes, and with the position and width of the RRL instability strip. 
Additionally, we inspect the database and identify a subsample of stars with relatively high iron abundances (${\rm [Fe/H]}>-0.5$\,dex).
In Section~\ref{sec:data}, we introduce the reader to the DESI survey, with an emphasis on its Y1 RRL catalog.
Section~\ref{sec:Bailey} describes the correlations between [Fe/H] and the periods and amplitudes of RRLs visible in the data, addressing the so-called Oosterhoff dichotomy, and the iron abundance and origin of short-period RRab and RRc. 
The metallicity dependence of the periods of pulsation of double-mode RRLs is described in Section~\ref{sec:rrds}. 
Section~\ref{sec:IS} details the observed dependency of the shape and width of the instability strip with metallicity from our RRL sample, and contrasts them with those predicted by stellar evolution models. 
In Section~\ref{sec:youngAndMrich}, we study the metal-rich end of the DESI RRL metallicity distribution, and discuss possible explanations for the stars in this regime.  
Lastly, we summarize our findings in Section~\ref{sec:conclusions}.

\section{Data}
\label{sec:data}

\subsection{The DESI RRL catalog}

The DESI Milky Way Survey (DESI-MWS) targets RRLs from the {\it Gaia} Data Release 2 \citep[DR2;][]{Clementini2019} and the Pan-STARRS1 \citep[PS-1;][]{Sesar2017} catalogs.
These RRLs were observed as primary and secondary targets in the commissioning, science verification, and main component of DESI, primarily as part of its bright and dark programs, with median exposure times of $\sim 750$ and $\sim1100$\,s (respectively). 
An overview of the DESI survey, its observing strategy, and data processing pipelines are provided by 
\citet{DESI_spectroscopic_pipeline_2023},\citet{DESI_Survey_Operation_2023},
\citet{DESI_Corrector_2024}
and \citet{DESI_Fiber_System_2024}.
For a comprehensive description of DESI-MWS, including details of its data products and its stellar catalog, we refer the reader to \citet{Cooper2023},  \citet{Koposov2024}, and \citet{Koposov2025}.
Further details of the RRL catalog, including a validation of the spectroscopic properties discussed in this paper, 
are presented by M25.

The RRL catalog used in this work was compiled by M25, taking advantage of DESI's high observing efficiency, wide wavelength coverage, and the large footprint of the survey. 
This catalog is based on observations taken by the DESI survey during its first year of operation (Y1), which correspond to 
DESI's first data release \citep[DR1;][]{DESI_DR1_2025}.
The catalog was assembled by crossmatching the DESI database with the {\it Gaia} DR3 RRL catalog \citep[][]{Clementini2023} and is composed of 6,240 RRLs, representing one of the largest catalogs with homogeneously-derived spectroscopic properties available to date (other surveys targeting large numbers of RRLs include LAMOST, GALAH, and APOGEE;  \citealt{Deng2012,DeSilva2015,Majewski2017}). 
The catalog includes spectroscopic properties corrected by the pulsation component of each RRL (one of the biggest challenges in the study of the spectra of RRLs), 
including systemic (i.e., center-of-mass) velocities, effective temperatures (\teff), surface gravities ($\log g$), and chemical abundances ([Fe/H] and [$\alpha$/Fe]). 
For a detailed description of the methodology used to correct the line-of-sight velocities and \teff\ of the sample for their expected variations throughout the RRL pulsation cycle, which can easily reach $\sim60$\,km\,s$^{-1}$ and 1,000\,K for a given RRL (respectively), we refer the reader to M25. 
For DESI, velocities and atmospheric parameters are computed using DESI's data processing pipelines, namely RVS \citep[][]{Koposov2019} (used for single epoch and coadded spectra) and SP \citep[][]{AllendePrieto2006} (for coadded spectra). 
Moreover, the DESI Y1 RRL catalog includes [Fe/H] derived using the recent calibration of so-called $\Delta$S method \citep[][]{Preston1959} presented by \citet{Crestani2021a}, which relates the equivalent widths of Balmer lines and the Ca K line with the [Fe/H] of RRLs. 
In this work, we rely mostly on the [Fe/H] from RVS, which are derived using the Python package RVSpecfit \citep[][]{Koposov:2019} based on the fitting of interpolated synthetic spectra from PHOENIX  model atmospheres \citep[][]{Husser2013}. 
A Bayesian inference methodology is then employed to estimate the systemic velocities and mean \teff\ of the RRLs, from which the variation of these properties throughout the RRL pulsation cycle is modeled.
This enables the estimation of these properties even from single-epoch measurements. 

The best classification of the RRLs in the DESI Y1 sample and their
overall variability properties (e.g., periods and amplitudes of pulsation), 
are taken directly from the {\it Gaia} DR3 catalog.
Based on this classification, 4,524 (72\%) RRLs in our sample are fundamental mode pulsators, 1,609 (26\%) are first-overtone pulsators, and 107 (2\%) are double mode pulsators.

The DESI Y1 catalog contains RRLs with heliocentric distances ($d_{\rm H}$) ranging between 2 and 115\,kpc.
In terms of Galactocentric distances $R_{\rm GC}$, calculated adopting a spheroidal halo and a distance of 8.275\,kpc from the Sun to the Galactic center \citep[][]{Gravity2021}, the sample covers a range of 3 to 120\,kpc. 
Thus, this catalog reaches regions well beyond the inner halo and with a significant coverage of the outer halo.
These RRLs span a wide range of iron abundances, from 
$-3.70$ to $-0.10$\,dex. 
 
A significant fraction of the RRLs in our catalog can be associated with known stellar systems and substructures.
Indeed, M25 determined that 600 are likely part of the Sagittarius (Sgr) stream, 2,134 have velocities and [Fe/H] consistent with the Gaia-Sausage-Enceladus (GSE) merger progenitor, and 86 are high probability members of dwarf spheroidal galaxies or globular clusters. 
More specifically, the DESI Y1 RRL sample contains RRLs in Draco (59 RRLs), Ursa Major~II (2), Sextans (2), Bootes~III (1), Palomar~5 (5), NGC~5904 (4), NGC~5466 (3), NGC~5024 (3), NGC~5053 (2), NGC~6341 (2), NGC~2419 (1), NGC~4147 (1), and NGC~5634 (1).

\subsection{Orbital parameters}

To determine the dynamical properties of the RRLs in our sample (which are relevant for Sections~\ref{sec:Bailey} and \ref{sec:youngAndMrich}), 
we model their orbits using the Python module {\sc GALPY} \citep[][]{Bovy2015}\footnote{ \href{http://github.com/jobovy/galpy}{http://github.com/jobovy/galpy}} 
integrating the orbits for 5\,Gyr backwards in time. 
To model the Galactic potential, we adopt {\sc GALPY}'s built-in {\it MWPotential2014}, which consists of a spherical nucleus and bulge \citep[Hernquist potential;][]{Hernquist1990}, a Miyamoto-Nagai disk model \citep[][]{MiyamotoNagai1975}, and a spherical Navarro-Frenk-White dark matter halo \citep[][]{Navarro1997}.
Additionally, we considered the impact of the LMC on the Milky Way potential, given the growing evidence that indicate its significant effects for modeling orbits \citep[see e.g.,][]{Erkal2018,Cunningham2020,Vasiliev2021,Medina2023}. 
For the current position of the LMC, we used $\alpha=78.77$\,deg and $\delta=-69.01$\,deg as central equatorial coordinates, and a distance of $d_{\rm LMC}=49.6$\,kpc \citep[][]{Pietrzynski2019}. 
For its current motion, we adopted a systemic line-of sight velocity of $262.2$\,km\,s$^{-1}$ \citep[][]{vanderMarel2002} and proper motions of $\mu_\alpha=1.850$\,mas\,yr$^{-1}$ and $\mu_\delta= 0.234$\,mas\,yr$^{-1}$ \citep[][]{GaiaCollaboration2018}.
We assumed an LMC mass of $1.88\times10^{11}$\,M$_\odot$ \citep[][]{Shipp2021} and a scale length $a_{\rm LMC}=20.22$\,kpc, which recovers the circular velocity at $8.7$\,kpc from the LMC center observed by \citet{vanderMarel2014}.  
To account for the fact that the LMC is not bound to the Milky Way in {\it MWPotential2014}, we multiply the Milky Way halo mass by a factor of 1.5 (which makes the LMC bound to the Galaxy\footnote{ \href{https://docs.galpy.org/en/stable/orbit.html}{https://docs.galpy.org/en/stable/orbit.html}}). 
Lastly, we consider the effect of Chandrasekhar dynamical friction in the LMC's orbit integration, and ignore the gravitational perturbations caused by other massive Milky Way satellites.

The orbital parameters of the RRLs in our sample and their uncertainties are obtained using {\it Gaia} DR3 proper motions and DESI-derived systemic velocities and heliocentric distances (M25).
We perform 500 {\sc GALPY} realizations per star, where each realization is obtained by randomly selecting a systemic velocity, heliocentric distance, and proper motions assuming that they follow a multi-dimensional Gaussian centered at the observed values, with standard deviations equal to their uncertainties. 
Proper motion uncertainties and their covariances are obtained directly from {\it Gaia} DR3, whereas systemic velocity and heliocentric distance uncertainties are derived from DESI's radial velocity curve modeling and error propagation, respectively (see M25).
The orbital parameters and their uncertainties correspond to the median of the resulting distributions and their 16 and 84 percentiles, respectively.

\section{The Bailey diagram}
\label{sec:Bailey}

The position of RRLs in the period-amplitude space, the so-called Bailey diagram, offers a valuable perspective to study the physics of their pulsation and the formation of the halo \citep[see e.g.,][]{Fiorentino2015,Fiorentino2017,Belokurov2018rrls,Fabrizio2019,Bono2020}. 
In this diagram, first-overtone pulsators are clumped in its short-period, low-amplitude region, whareas RRab variables are characterized by larger amplitudes and longer periods than RRcs (overall), with luminosity amplitudes decreasing with increasing periods.
The distribution of RRLs in the Bailey diagram depends on their physical properties (evolutionary phase, mass, luminosity, effective temperature) and chemical composition, as well as the existence of secondary modulations and non-linear pulsation phenomena \citep[e.g., shocks;][]{Bono2020}.

\begin{figure}
\begin{center}
\includegraphics[angle=0,scale=.32]{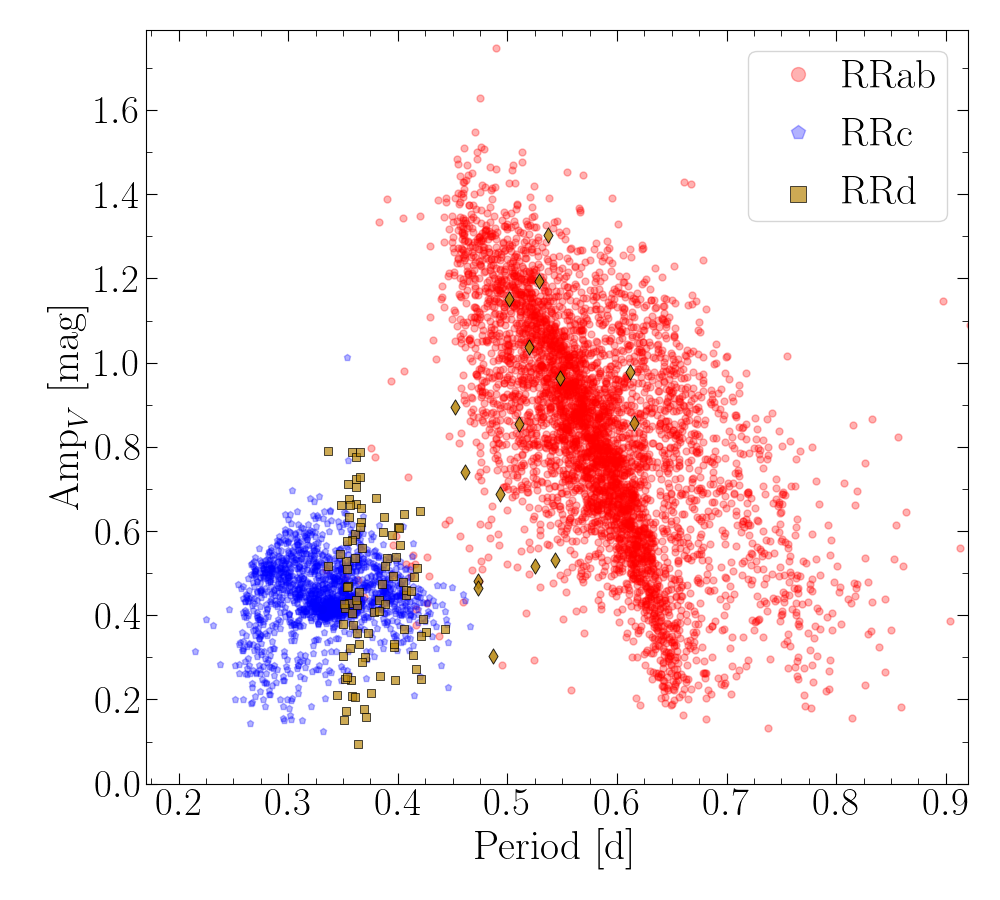}
\caption{
Bailey diagram of the DESI Y1 RRL sample (adapted from M25). 
The light curve amplitudes in the $V-$band (${\rm Amp}_V$)
shown are computed using the peak-to-peak $G-$band magnitudes listed in the {\it Gaia} catalog and the transformation equation from \citet{Clementini2019}. 
The classification of RRLs between RRab, RRc, and RRd is taken directly from the {\it Gaia} DR3 catalog \citep{Clementini2023}. 
Anomalous RRd stars are shown with diamond markers (see Section~\ref{sec:rrds} for more details about these stars). 
}
\label{fig:bailey}
\end{center}
\end{figure}

\begin{figure*}
\begin{center}
\includegraphics[angle=0,scale=.325]{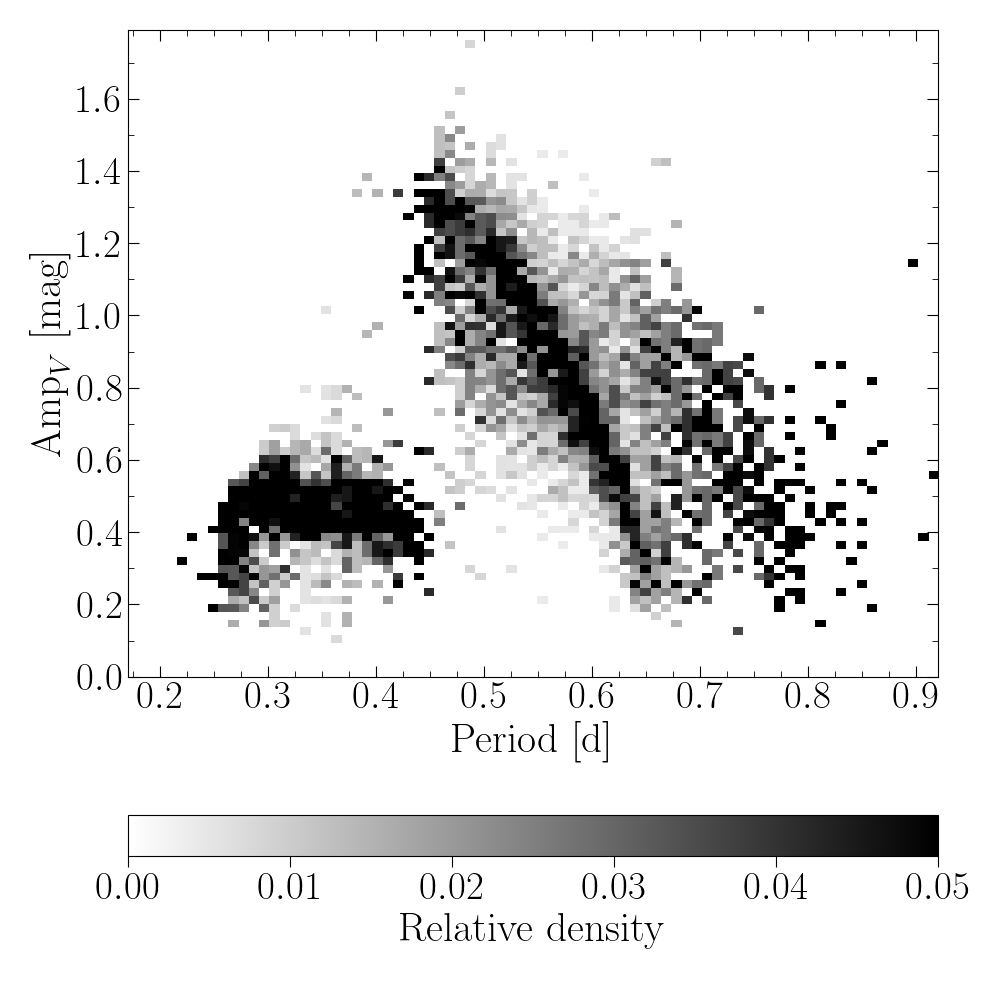}
\includegraphics[angle=0,scale=.320]{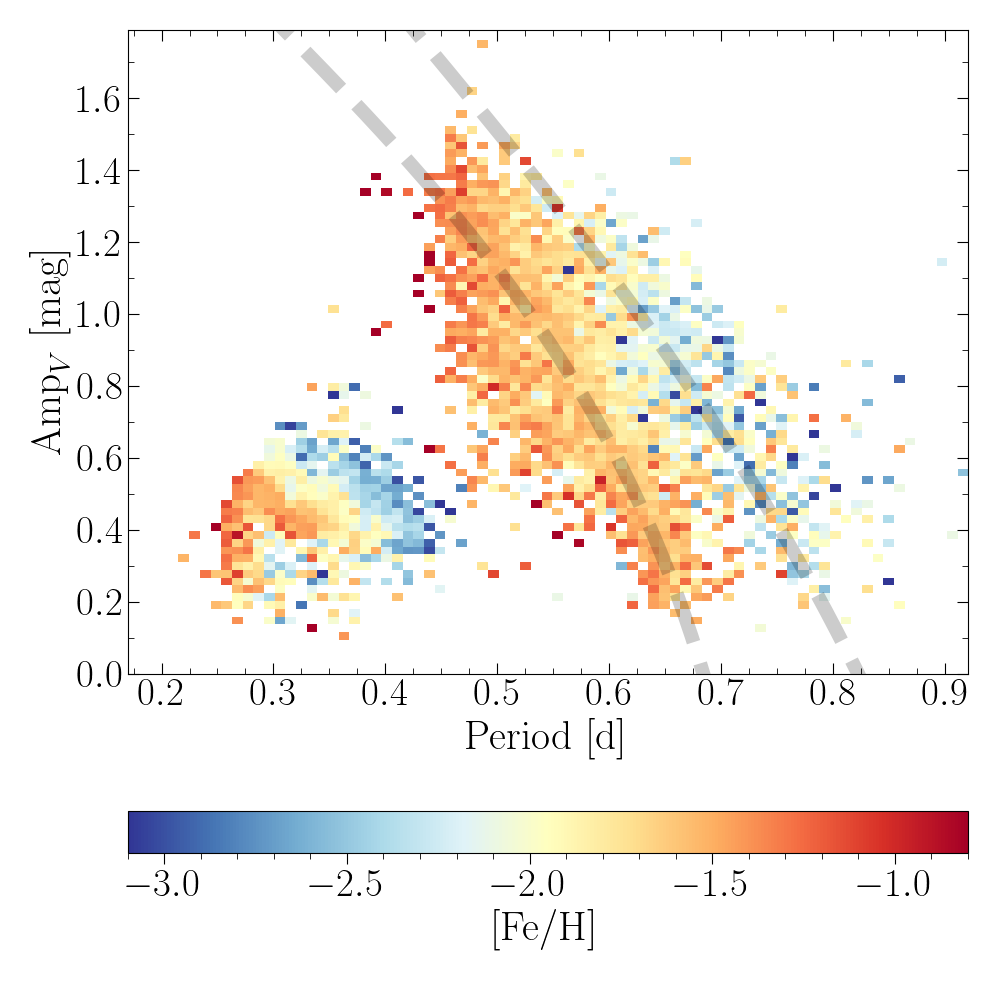}
\caption{
Same as Figure~\ref{fig:bailey} but as a column normalized histogram ({\it left}, i.e. the sum of the density in each column is one) and  color-coded by the average [Fe/H] in each pixel ({\it right}).
In the right plot, dashed lines are used to represent the position of the fiducial curves for OoI and OoII defined by \citet{Fabrizio2019}.
The position of the OoI and OoII lines is consistent with the location of the bulk of the more metal rich ([Fe/H] $> -2.0$\,dex) and more metal-poor ([Fe/H] $<-2.0$\,dex) RRab population in the Bailey diagram, respectively.
}
\label{fig:bailey_MET}
\end{center}
\end{figure*}

Figure~\ref{fig:bailey} depicts the Bailey diagram for the DESI Y1 sample for RRab, RRc, and RRd variables, using the periods provided in the {\it Gaia} DR3 catalog and $V-$band pulsation amplitudes (${\rm Amp}_V$).
For the latter, we adopt the prescription of \citet{Clementini2019} to transform the peak-to-peak amplitudes reported by {\it Gaia}, ${\rm Amp}_G$ ({\sc peak\_to\_peak\_g} in {\it Gaia}) to $V-$band amplitudes, i.e., ${\rm Amp}_V = (1.081 \pm 0.003)\ {\rm Amp}_G + (0.013 \pm 0.003)$.
Figure~\ref{fig:bailey} shows clearly the separation between the RRL subtypes, from their classification in the {\it Gaia} catalog. 
In the rest of this section, we analyze two science questions related to the position of the DESI Y1 RRL sample in the Bailey diagram.

\begin{figure*}
\begin{center}
\includegraphics[angle=0,scale=.40]{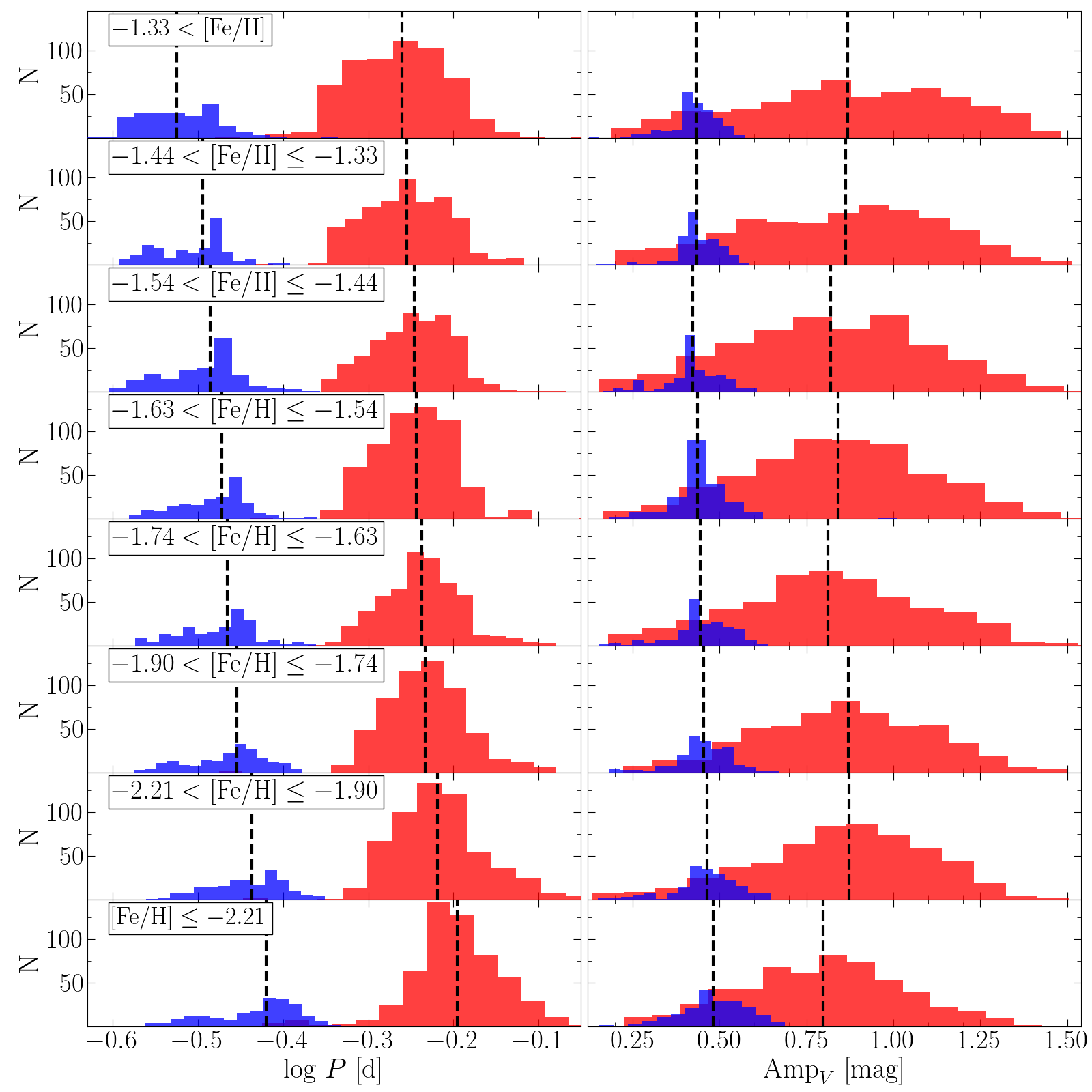}
\caption{
Distribution of log$P$ (left panels) and visual amplitude ${\rm Amp}_V$ (right panels) of the DESI Y1 RRL sample as a function of iron abundance. 
To illustrate existing correlations, the data is split into eight metallicity bins containing the same number of stars. 
In each panel, a blue histogram is used to represent RRc stars, whereas the log$P$ and ${\rm Amp}_V$ distribution of RRab is shown in red (towards longer periods and higher amplitudes). 
Dashed vertical lines display the median of each distribution. 
The left-hand panels show a systematic shift in log$P$ from the more metal-rich to the more metal-poor regime towards long periods, both for RRab and RRc. 
This trend is more subtle for the visual amplitudes of RRc and not observed for RRab. 
}
\label{fig:perampmet}
\end{center}
\end{figure*}

\subsection{On the Oosterhoff dichotomy and [Fe/H] dependencies in the Bailey diagram}
\label{sec:OoAndBailey}

One of the most intriguing and long-standing problems related to the position of RRLs in the Bailey diagram is the so-called Oosterhoff dichotomy \citep[see e.g.,][]{Caputo1989,Bono1994_Oosterhoff,Bono1995_Oosterhoff,Lee1999,Cassisi2004}.
The Oosterhoff dichotomy was first reported by \citet{Oosterhoff1939}, who recognized that RRLs in globular clusters could be mainly separated into two groups: 
one with RRab variables with periods $\leq0.55$\,d (Oosterhoff I, or OoI), and the other with RRab periods closer to 0.65 (Oosterhoff II, or OoII).  
It was later recognized that this dichotomy is also reflected in the fraction of RRc with respect to the total number of RRLs in the clusters, with a larger fraction corresponding to OoII \citep[see e.g.,][]{Braga2016}. 
Furthermore, OoI clusters have been found to be more metal-rich than those in the OoII group ([Fe/H] $\sim-1.3$ vs. $\sim-2.0$\,dex, respectively; see e.g., Section~6 in \citealt{Catelan2009}, and \citealt{Catelan2015}). 
On the other hand, RRab stars in dwarf galaxies tend to populate the period gap between the two Oosterhoff groups (the {\it Oosterhoff gap}), in the so-called Oosterhoff-intermediate (Oo-int) region, and field stars are predominantly OoI (see  \citealt{Catelan2015} and references therein).
Thus, the fact that RRab in dwarf galaxies and in the field do not follow the dichotomy, together with findings of dissimilar chemical enrichment histories between globular clusters and the Milky Way halo \citep[see e.g.,][]{Carretta2009}, 
gives us insight into the role of globular clusters in the formation of the Galactic halo.

The homogeneous dataset provided by DESI 
offers a unique opportunity to inspect metallicity variations in the Bailey diagram and to study their association with the Oosterhoff dichotomy.
For this analysis, we employ the iron abundances obtained with DESI's RVS processing pipeline. 
The left panel of Figure~\ref{fig:bailey_MET} depicts the Bailey diagram of the DESI Y1 RRL sample as a column-normalized histogram, showing two distinctive groups of RRab stars associated with the two Oosterhoff groups. 
The right panel of Figure~\ref{fig:bailey_MET} shows the correlation of the position of the RRL in the period-${\rm Amp}_V$ with their iron abundance. 
In the right panel, the fiducial OoI and OoII lines defined by \citet{Fabrizio2019} using a similarly large dataset (for RRab stars) are included, showing an overall good agreement with our data.
We note in passing that our sample is composed predominantly of field stars, of which a significant fraction exhibits chemodynamics consistent with the GSE accretion event. 
These stars are mainly associated with the OoI group. 
Among the RRab in intact satellites, those in Draco are the most numerous (46 stars),  and exhibit a mean period of $0.60$\,d with a standard deviation of $0.07$\,d.
The fundamental-mode RRLs (RRab) in GCs are less numerous (16 stars), but seem to be consistent with the dichotomy with a preferential location towards the locus of the OoII group (as shown later in Figure~\ref{fig:baileyhasps}), with a mean period of their RRab of $0.64$\,d, and standard deviation of $0.08$\,d. 
Similar to the RRab in Draco, those in the Sgr stream (465 RRab stars) display a mean period of $0.59$\,d with a standard deviation of $0.06$\,d, broadly consistent with the OoI group.

In Figure~\ref{fig:bailey_MET}, we observe a smooth decline in metallicity (from the metal-rich to the metal-poor end of the RRL distribution) as the period of pulsation increases, at any given ${\rm Amp}_V$ and for both RRab and RRc. 
For a direct comparison with the period and ${\rm Amp}_V$ [Fe/H] correlations recently reported by  \citet{Fabrizio2021}, we first follow their methodology and split our RRab and RRc samples in metallicity bins, each containing the same number of stars. 
Here, however, we use eight bins per sub-class to ensure that we obtain a number of stars per bin similar to those of \citet{Fabrizio2021} 
(566 RRab and 202 RRc per metallicity bin).

Figure~\ref{fig:perampmet} displays the overall trends that the distribution of periods (left panels) and visual amplitudes (right panels), for each sub-class, follow as a function of [Fe/H]. 
The figure shows a steady decrease in period when moving from the metal-poor to the metal-rich regime, for both RRab and RRc stars. 
In fact, the median of the RRab and RRc log$P$ distributions (shown as vertical dashed lines in Figure~\ref{fig:perampmet}) vary from log$P=-0.20$\,d to log$P=-0.26$\,d and from log$P=-0.42$ to log$P=-0.53$, respectively, when the [Fe/H] varies from the most metal-poor to the metal-rich bins.  
These trends are more subtle in the ${\rm Amp}_V$ panels (right-hand plots) for RRc stars and no clear trend is visible for RRab stars.

\begin{figure}
\begin{center}
\includegraphics[angle=0,scale=.345]{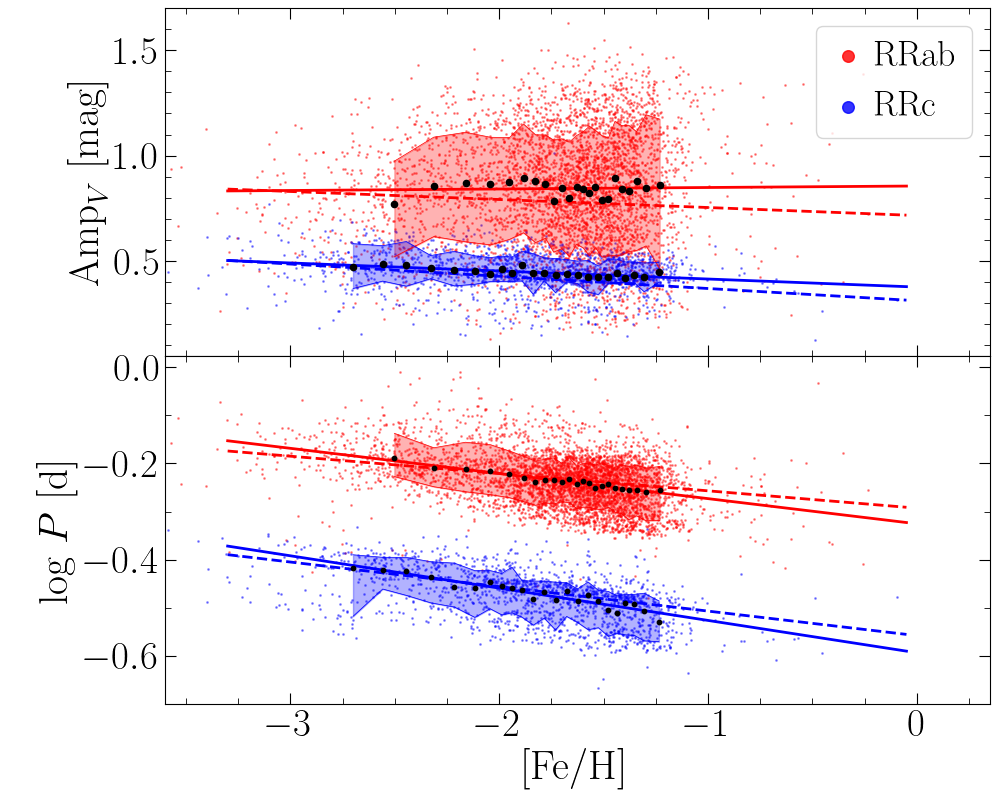}
\caption{ 
Iron-abundance dependence of the logarithm of the period and visual amplitude for RRab (red) and RRc (blue). 
The data is split into [Fe/H] bins with equal number of stars, in the range [$-2.60$, $-1.19$]\,dex for RRab stars, and [$-2.79$, $-1.19$]\,dex for RRc stars. 
In each panel, black filled circles are used to represent the median of the log$P$ or ${\rm Amp}_V$ distribution in each bin.
Solid lines depict the best fit linear correlations for the medians computed per bin. 
Shaded regions show the 16th and 84th percentiles of each [Fe/H] bin. 
Dashed lines represent the ${\rm Amp}_V$-[Fe/H] and log$P$-[Fe/H] correlations reported by \citet{Fabrizio2021} using high-resolution and low-resolution ($\Delta$S method) spectra.
}
\label{fig:perampmetcorrelation}
\end{center}
\end{figure}

\begin{figure}
\begin{center}
\includegraphics[angle=0,scale=.34]{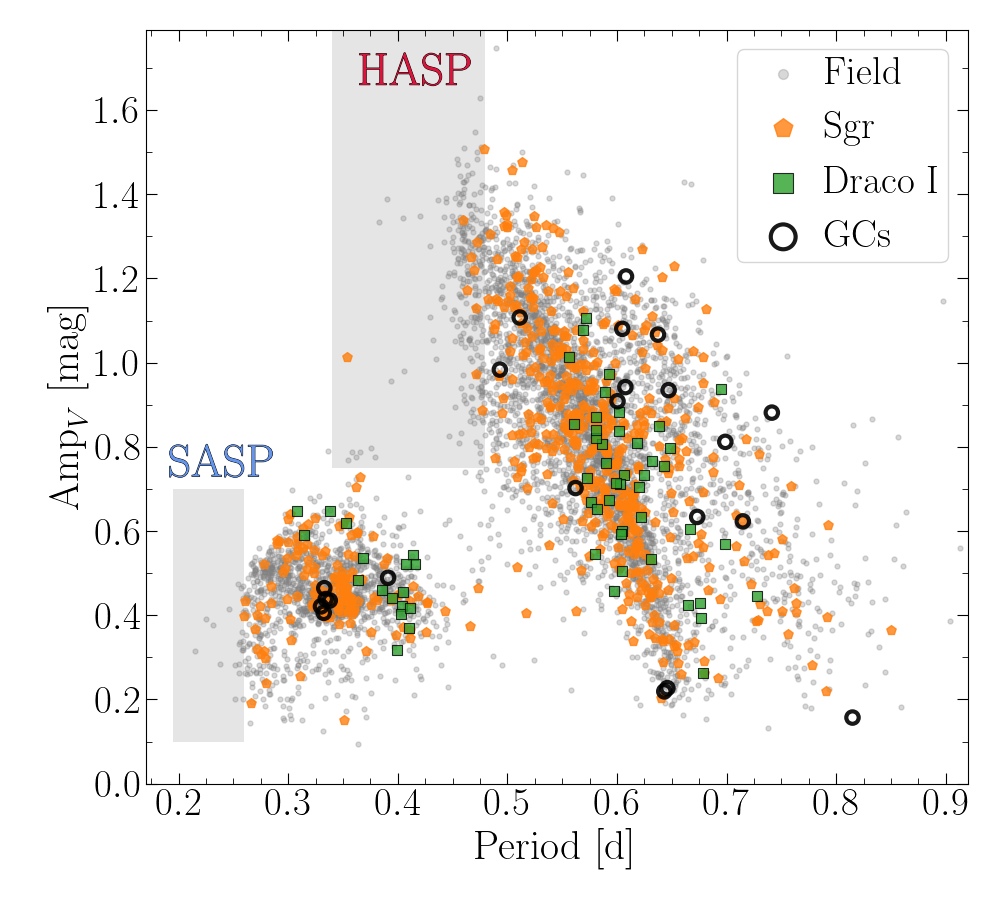}
\caption{
Bailey diagram of the DESI Y1 RRL sample separated by system.
This distinction is made to better illustrate the distribution of RRLs in the halo, the Sagittarius stream (Sgr), Draco, and the combined sample of globular cluster RRLs. 
We note that the HASP region is dominated by field stars (with a few stars from Sgr), and the SASP region contains only field stars.
This is in agreement with the general expectation for the origin of halo stars (see the text for more details). 
}
\label{fig:baileyhasps}
\end{center}
\end{figure}

In order to better quantify the [Fe/H] dependence of the periods and amplitudes, we repeat the experiment increasing the number of metallicity bins and using an equal number of stars per bin. 
We split the sample in 25 [Fe/H] bins for RRab and RRc stars, each containing 181 and 65 stars, respectively. 
Then, we compute the median of the log$P$ and ${\rm Amp}_V$ of the stars in each bin, as well as their 16th and 84th percentiles.
Figure~\ref{fig:perampmetcorrelation} depicts the log$P$ and ${\rm Amp}_V$ of the entire sample as a function of [Fe/H], as well as the median and percentiles in each bin, as described above. 
The figure shows that both visual amplitudes and periods of RRc stars vary smoothly over the studied metallicity range, while for RRab stars only the periods exhibit a noticeable change. 
We use these data to fit linear relations between the parameters with the \code{optimize} module of the Python package \code{scipy}, in the [Fe/H] range [$-2.60$,$-1.19$]\,dex for RRab stars, and [$-2.79$,$-1.19$]\,dex for RRc stars. 
It is noteworthy that such linear correlations are predicted by the pulsation relation when the evolutionary and pulsation properties of RRLs are taken into account \citep[][]{vanAlbada1973,Caputo1998}. 
The best fits for the correlations of each subtype are shown in Figure~\ref{fig:perampmetcorrelation} with solid lines, and correspond to:

\begin{equation}
\begin{array}{ll}
{\rm Amp}_V & = 0.856\ (\pm 0.041) + 0.007\ (\pm 0.024)\ {\rm [Fe/H]},\\
\log P & = -0.326\ (\pm 0.005) - 0.052\ (\pm 0.003)\ \rm{[Fe/H]} 
\end{array}
\label{eq:AVPERfeh_relation_rrab}
\end{equation}

\noindent for RRab stars and 

\begin{equation}
\begin{array}{ll}
{\rm Amp}_V & = 0.377\ (\pm 0.012) - 0.052\ (\pm 0.003)\ {\rm [Fe/H]},\\
\log P & = -0.594\ (\pm 0.010) - 0.067\ (\pm 0.005)\ \rm{[Fe/H]} 
\end{array}
\label{eq:AVPERfeh_relation_rrc}
\end{equation}

\noindent for RRc stars. 
These empirical relations indicate that an increase of 1\,dex in [Fe/H] results in an increase of $\sim 0.01$ in amplitude for RRab stars, and a decrease of $\sim 0.05$\,mag for RRc stars.
Similarly, an increase of 1\,dex in [Fe/H] yields $\sim 0.05$ and $\sim 0.07$\,dex lower values in the logarithm of the period of RRab and RRc stars, respectively. 
We observe that the coefficients used for the metallicity dependence of the periods are roughly consistent between RRab and RRc stars (both being negative and statistically different than zero). 
A quick inspection of the figure shows, however, that although the visual amplitude and the metallicity of RRab stars can be written as a rough correlation, the slope of the correlation is not significantly different from zero.
In that regard, we note that RRab stars display a broad distribution of ${\rm Amp}_V$ values in each bin (as also seen in Figures~\ref{fig:bailey} and \ref{fig:bailey_MET}) and large intervals computed from the 16th and 84th percentiles for these stars (of the order of $\sim 0.28$\,mag), in comparison to those in the ${\rm Amp}_V-$[Fe/H] space of the RRc stars ($\sim0.16$\,mag). 
We consider the different behavior for the two types of RRLs significant regardless of the inherently different ${\rm Amp}_V$ for RRab and RRc stars.
A tentative explanation for this difference is the impact of non-linear phenomena being stronger on fundamental-mode pulsators than on first-overtones \citep[see e.g.,][]{Sneden2017,Gillet2019}.  

Our derived ${\rm Amp}_V$ and log$P$ anticorrelations with [Fe/H] can be compared with those found by \citet{Fabrizio2021} from a similarly large dataset.
For RRab stars, we find almost no dependency of ${\rm Amp}_V$ on [Fe/H] (vs. a slope of $-0.038$\,mag\,dex$^{-1}$ in their work)
and a stronger dependency in the case of log$P$ (by a factor of 1.4).
Interestingly, for RRc variables the [Fe/H]-dependent factor on visual amplitude is just mildly smaller (by a factor of 1.1), but we find a similar difference in the logarithmic period as with the RRabs (i.e., by a factor of 1.3).
A potential explanation for the observed differences is that, although both of our works transform {\it Gaia} $G-$band amplitudes to the $V$ band for deriving the correlations, we differ in the method used to estimate the $G$ amplitudes, in the first place. 
More specifically, we use {\sc peak\_to\_peak} {\it Gaia} amplitudes whereas \citet{Fabrizio2021} adopt a combination of {\it Gaia} DR2-based analytical fit amplitudes and literature ${\rm Amp}_V$. 
Moreover, we follow different methodologies to derive [Fe/H]. 
When using DESI's $\Delta$S-based iron abundances (instead of the DESI spectroscopic metallicities we used for this part of the analysis), we find a much better agreement with their reported results for the zeropoints and the slopes of these correlations (within 1$\sigma$), with the exception of the zeropoint of our RRab ${\rm Amp}_V$-[Fe/H] relation (which differs from theirs by 3$\sigma$). 
This confirms the importance of the method for determining the [Fe/H] and $G-$band amplitude of the sample in the derived (empirical)  ${\rm Amp}_V$-[Fe/H] relation for RRab stars.

Several recent studies have confirmed the complex nature of the Oosterhoff dichotomy, as a result of stellar and galactic evolution processes. 
\citet{Zhang2023}, for instance, indicated that the properties that define the Oosterhoff dichotomy vary with the position in the Galaxy and from galaxy to galaxy. 
Other studies have directly attributed the emergence of the Oosterhoff dichotomy to the accretion history of the Galaxy and/or RRL evolutionary effects \citep[e.g.,][]{Luongo2024,Li2024}, while others have linked the depopulation of the Oosterhoff gap to the mass of the accreted dwarf galaxies hosting globular clusters \citep{Prudil2024c}.
\citet{Fabrizio2019} and \citet{Fabrizio2021} provided an interpretation for the observed dichotomy, based on their 
reported anti-correlation between the position of RRLs in the Bailey diagram and their metallicities (for both RRab and RRc), observed as a steady decrease in their pulsation periods and in their visual amplitudes with increasing metallicity.
These authors proposed the lack of metal-intermediate Galactic globular clusters hosting sizeable samples of RRL stars as a potential explanation for the dichotomy.  
Our reported anti-correlations (resulting from the extent of the instability strip exploration as a function of [Fe/H] and the period-density relation of RRLs) and the smooth transition from the long period (low metallicity) to the short period regime (high metallicity), together with the lack of intermediate metallicity Galactic globular clusters, provides a sensible explanation for the dichotomy, in line with the interpretation suggested by these authors.
We note, however, that while our results support a metallicity-driven interpretation of the Oosterhoff dichotomy, metallicity alone does not uniquely determine horizontal branch morphology (and therefore, the pulsation properties of RRLs). 
The well-known second parameter problem, i.e., the dependence of horizontal branch morphology on parameters other than metallicity (e.g., age, helium abundance, and mass loss), also plays a role in shaping the distribution of stars along the instability strip and the observed Oosterhoff properties \citep[see e.g.,][]{Catelan2009,Catelan2015}.

\subsection{HASPs and SASPs in DESI and the origin of the halo}
\label{sec:haspsasp}

As already discussed, the position of RRL variables in the Bailey diagram and their metallicities are important for their role as tracers of the halo formation and their connection with its building blocks. 
Indeed, several authors have shown that most Milky Way dwarf galaxy satellites lack high-amplitude short-period (HASP) RRab variables, that is, those with periods $\leq0.48$\,d and visual amplitudes $\geq0.75$\,mag, whereas they are typically observed in globular clusters (with [Fe/H] $\geq-1.4$\,dex) and in the halo field \citep[][]{Fiorentino2015,Fiorentino2017,Monelli2022}.
This can be interpreted as a consequence of the correlation between these quantities and [Fe/H], as the period distribution of RRab in relatively metal-rich globular clusters is shifted toward short periods \citep[see e.g.,][]{Drake2013,Fiorentino2015}. 
HASPs are also observed in the most massive Milky Way satellites (e.g., Sgr and the Magellanic Clouds), which suggests that the stars currently observed in most dwarf galaxies differ from the building blocks of the halo and that these building blocks were likely systems more massive than $10^9$\,M$_\odot$. 
In other words, the presence of HASPs in a system (determined solely by their position in the Bailey diagram) seems to be an indication of a fast early chemical enrichment, which allows the formation of old stars with [Fe/H] $\geq-1.4$\,dex.
For an updated census of HASPs in dwarf galaxies and globular clusters, we refer the reader to Figure 7 in \citet{Monelli2022}.

In a similar vein, small-amplitude short-period (SASP) RRc variables (those with periods $<0.26$\,d) are not present in dwarf galaxies \citep[][]{Fiorentino2022}, which also arises from the dependency between their periods and metal content.  
We note in passing that the existence of moderately metal-rich RRLs (with $-1.5\lesssim{\rm [Fe/H]}\lesssim-0.5$) is predicted by theoretical models, based on their pulsational and evolutionary properties \citep[see][]{Bono1997a,Bono1997b}.
In fact, similar to HASPs, the period distribution of SASPs is only shifted toward short periods for RRc stars in massive satellites and in the halo (which corresponds to [Fe/H] $\sim-1.4$\,dex; \citealt{Crestani2021b}), another indication of the minor contribution of systems less massive than $10^9$\,M$_\odot$ to the formation of the halo.

Figure~\ref{fig:baileyhasps} shows the DESI Y1 Bailey diagram highlighting the HASP and SASP regions, making the distinction between RRLs in the halo field, the Sgr stream, Draco, and the combined sample of globular cluster RRLs. 
Here, following the convention from the literature \citep[][]{Fiorentino2015,Fiorentino2017,Fiorentino2022}, 
we define the HASP region as that with periods $<0.48$\,d and ${\rm Amp}_V>0.75$\,mag, and the SASP region as that with periods $<0.26$\,d and ${\rm Amp}_V>0.1$\,mag.
It is clear from the figure that the HASP region is dominated by field stars.
In fact, of the 270 RRab stars in this region, 97\% are field stars, and only nine of them (3\%) are from the Sgr stream. 
Similarly, 22 out of the 23 RRc stars in the SASP region are field stars (96\%), and only one of them is from Sgr. 
For comparison, 88\% of the non-HASP RRab stars are field RRLs, whereas for non-SASP RRc stars this fraction is 91\%. 
Interestingly, $\sim60$\% of the RRab in the HASP region display GSE membership probability $p_{\rm GSE}>0.7$ (as reported by M25).
For SASP RRc stars, $\sim40$\%  exhibit $p_{\rm GSE}>0.7$.
There are no HASPs or SASPs in Draco and in globular clusters, following the expected trend\footnote{To put this statement into context, we note that all globular clusters hosting RRLs in our sample have ${\rm [Fe/H]}<-1.5$, with the exception of Pal~5 and NGC 5904 (for both of which ${\rm [Fe/H]}\sim-1.3$; \citealt{Bailin2022}).}. 

\begin{figure}
\begin{center}
\includegraphics[angle=0,scale=.33]{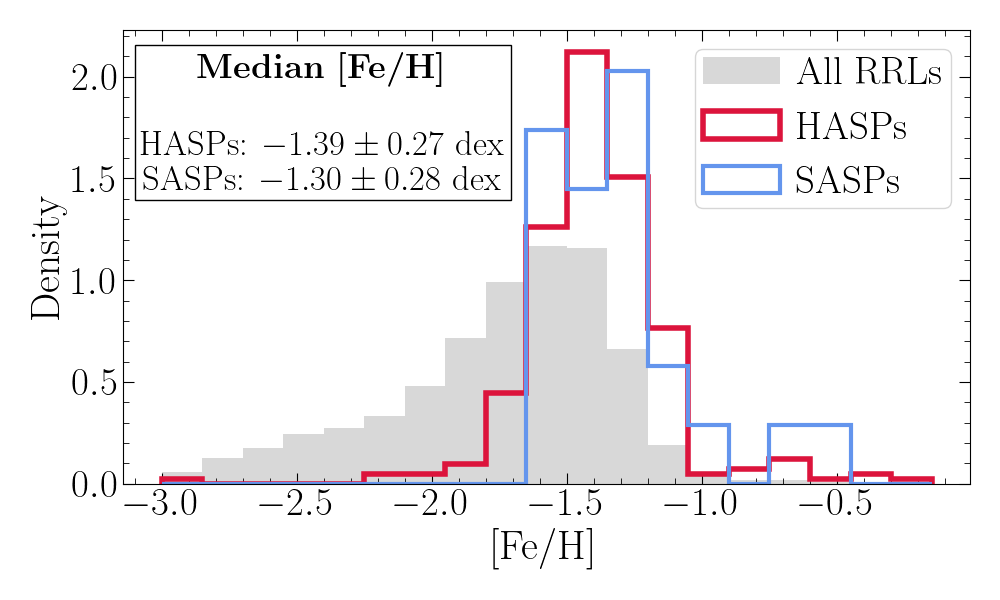}
\includegraphics[angle=0,scale=.33]{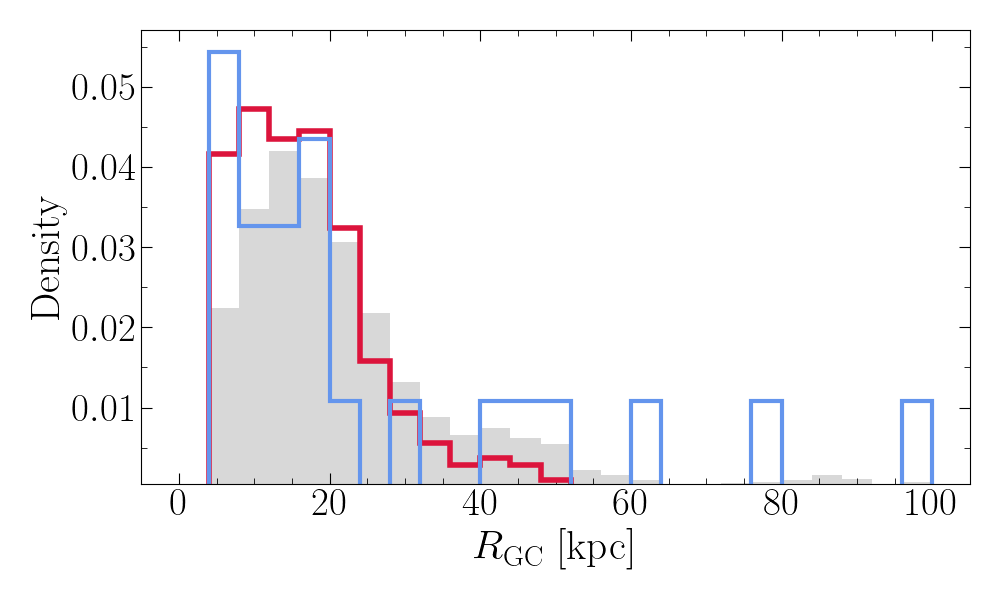}
\includegraphics[angle=0,scale=.33]{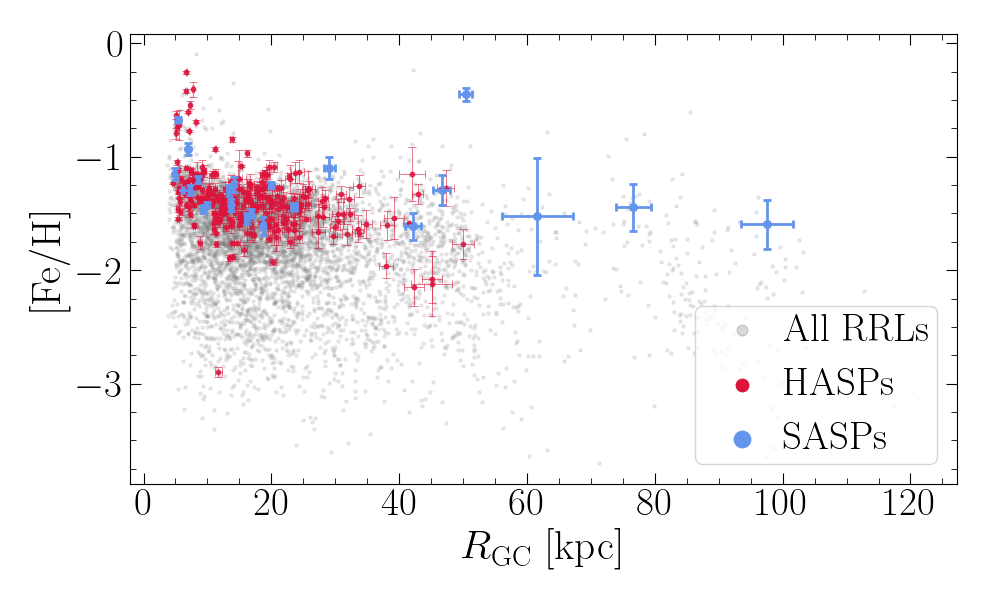}
\caption{
Normalized [Fe/H] (top) and Galactocentric distance (center) histograms in the DESI Y1 RRL sample}, highlighting the distribution of HASPs (red) and SASPs (blue). 
All histograms are normalized as probability densities (i.e., the area under each histogram equals unity). 
The bottom panel displays the metallicity as a function of $R_{\rm GC}$ for these samples. 
\label{fig:metdist_hasps_sasps}
\end{center}
\end{figure}

To quantify the iron content distribution of short period RRLs, Figure~\ref{fig:metdist_hasps_sasps} depicts [Fe/H] for our sample of HASPs and SASPs. 
As shown in the figure, the bulk of the metallicity distribution of both groups lies at $-2.0\leq {\rm [Fe/H]} \leq-1.0$\,dex, with a few outliers in both ends of the distribution.
The median metallicity corresponds to $-1.39$\,dex and $-1.30$\,dex for HASPs and SASPs, with standard deviations of $0.27$\,dex and $0.28$\,dex, respectively. 
Thus, our results are consistent with previous theoretical and empirical predictions for short-period RRLs.

\begin{figure}
\begin{center}
\includegraphics[angle=0,scale=.43]{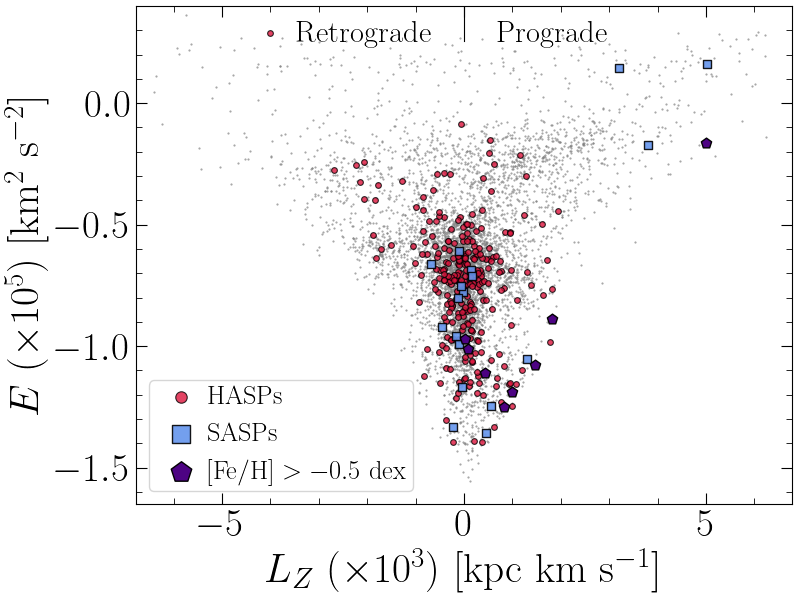}
\caption{
Vertical angular momentum $L_Z$ and energy $E$ of the stars in the DESI Y1 RRL sample, computed from the integration of their orbits using {\sc GALPY} \citep[][]{Bovy2015}. 
Red, blue, and purple markers represent HASPs, SASPs, and RRLs with [Fe/H]$>0.5$\,dex, respectively, whereas stars that do not fall in these categories are shown with grey dots. 
}
\label{fig:LzvsE_HASP_SASP_MRYR}
\end{center}
\end{figure}

Figure~\ref{fig:metdist_hasps_sasps} also displays the distance distribution of the HASP and SASP samples. 
A marked decline in the distance distribution of these stars beyond $\sim20$\,kpc is clear from the figure (similar to the decline of the full distribution, also shown in the figure).
This decline is specially clear for HASPs, which are more abundant than SASPs in our sample. 
However, we find a non-negligible fraction of these samples beyond this limit (79 HASPs and 8 SASPs), and even beyond 30 kpc (23 HASPs and 6 SASPs). 
The outer halo of the Milky Way is predominantly composed of stars originating from the accretion of satellites \citep[see e.g.,][]{Naidu2020,Medina2024}.
Since stars are not expected to form in-situ at these distances, and the aforementioned properties are suggestive of the nature of their origins, this sample displays strong evidence of having been accreted from massive satellites.

A group of HASPs and SASPs with ${\rm [Fe/H]}>-1$\,dex is observed at $R_{\rm GC}<20$\,kpc.
These stars are characterized by prograde motions and disk-like kinematics (based on their energies and angular momenta, as described later in this section).
The figure also shows that our SASP sample reaches $R_{\rm GC}\sim100$\,kpc, whereas the HASP stars are only found out to $R_{\rm GC}\sim50$\,kpc.
A possible explanation for the presence of SASPs beyond 50\,kpc (and for the unexpectedly high [Fe/H] of a SASP at $\sim 50$\,kpc) is the contamination in the {\it Gaia} DR3 RRL catalog at large distances (e.g., by eclipsing binary stars). 
This suggestion is supported by the low score of the best class (RRc) reported by {\it Gaia} for these stars ({\sc best\_class\_score} $<0.1$ for all of them). 
In this context, a systematic examination of the RRL light curves (e.g., periods, amplitudes, Fourier parameters, and visual inspection of phased light curves) would therefore help identify and remove clearly misclassified variables before drawing firm conclusions.
Additionally, the figure shows an outlier in the HASP metallicity distribution at $R_{\rm GC}\sim10$\,kpc, for which ${\rm [Fe/H]}\sim-2.9$\,dex ($\sim-3.3$\,dex if measured by DESI's SP pipeline, and $\sim-1.5$\,dex from the $\Delta$S method) with high signal-to-noise ratio S/N ($\sim65$) and relatively high {\sc best\_class\_score} (0.82).
This star, however, is flagged as an unsuccessful model fit by DESI's SP pipeline likely due to data reduction. 
Lastly, we note that SASP stars and the bulk of the HASP sample follow the expected metallicity gradient of the DESI RRLs in GSE (with a slope of $0.005$\,dex\,kpc$^{-1}$; Figure~15 in M25), with larger scatter around this trend for HASPs at large distances ($R_{\rm GC}\geq35$\,kpc) likely due to number statistics. 
This supports their potential association with this merger event.   
Provided that these stars are bona-fide RRLs, we advocate for detailed high-resolution spectroscopic follow-up studies of the most distant HASPs and SASPs to confirm or reject their association with a massive progenitor via the expected high [Fe/H] observed for these stars as well as other chemical-abundance-based signatures, such as [$\alpha$/Fe] (e.g., by inspecting sequences extending from the metal-poor $\alpha$-enhanced regime to solar metallicities depleted in $\alpha$ elements).

Figure~\ref{fig:LzvsE_HASP_SASP_MRYR} depicts the energy ($E$) and vertical angular momentum ($L_Z$) of our sample, highlighting the position of HASPs and SASPs in this space. 
The relative position of stars in this diagram serves as an indication of their dynamical origin, as stars with a common progenitor share similar integrals-of-motion due to the long timescales required for phase-space mixing in the halo. 
From their $E$-$L_Z$ distribution, it is clear that a significant fraction of these stars lie close to $L_Z\sim0$\,kpc\,km\,s$^{-1}$, a region dominated by stars with radial orbits associated with the GSE merger event (as shown by M25). 
Together with the fraction of HASP and SASP stars with high $p_{\rm GSE}$ and the observed [Fe/H] as a function of $R_{\rm GC}$ (Figure~\ref{fig:metdist_hasps_sasps}), this supports the association of these stars with GSE. 
For the non-GSE HASPs, we observe a relatively homogeneous $E$ and $L_Z$ distribution, with $E<0$\,km$^{2}$\,s$^{-2}$ and displaying both prograde ($L_Z>0$\,kpc\,km\,s$^{-1}$) and retrograde ($L_Z<0$\,kpc\,km\,s$^{-1}$) orbits. 
Interestingly, this is not the case for our SASP sample, which show orbits that are preferentially prograde and relatively low energy ($E<-0.5\times10^{5}$\,km$^{2}$\,s$^{-2}$; see Figure~\ref{fig:LzvsE_HASP_SASP_MRYR}), with the exception of three stars, all of which lie at $R_{\rm GC}>30$\,kpc (two of them with {\sc best\_class\_score} $<0.1$).

\section{Double-mode pulsators and [Fe/H]}
\label{sec:rrds}

The ratio of the first-overtone period to the fundamental mode period of RRd stars provides reddening independent insights on the physical properties of these stars. 
Indeed, the position of RRd stars in the first-overtone to fundamental period ratio ($P_{\rm 1O}/P_{\rm F}$) versus fundamental mode period ($P_{\rm F}$) space, the so-called Petersen diagram \citep[][]{Petersen1973}, has been extensively studied both empirically \citep[see e.g.,][]{Soszynski2019,Chen2023,Braga2022} and theoretically \citep[with the use of pulsation models; see e.g.,][]{Cox1980,Bono1996,Popielski2000}.  
In the Petersen diagram, RRd stars follow a well-defined sequence, as a consequence of the impact of luminosity, stellar mass, \teff, and chemical composition on these stars' periods \citep[stemming from the van Albada and Baker relations;][]{vanAlbada1971}.
Furthermore, ``anomalous'' RRd stars have been classified based on their position in the Petersen diagram, where they lie above or below the locus of the period ratio-$P_{\rm F}$ distribution of ``classical'' RRd stars. 
These stars have been detected in multiple environments \citep[][]{Jurcsik2015,Soszynski2016} and appear to be less common than classical RRd stars, overall. 
Anomalous RRd stars often display indications of long-term amplitude, period, and phase modulation \citep[see e.g.,][]{Soszynski2014, Smolec2015, Netzel2022}, the so-called Blazhko effect \citep[][]{Blazhko1907}\footnote{The Blazhko effect was originally observed in RRab stars, and later in RRc  stars, before found in
anomalous RRd variables \citep[see e.g.,][and references therein]{Catelan2015}.}, and are characterized by dominant fundamental modes.

\begin{figure}
\begin{center}
\includegraphics[angle=0,scale=.415]{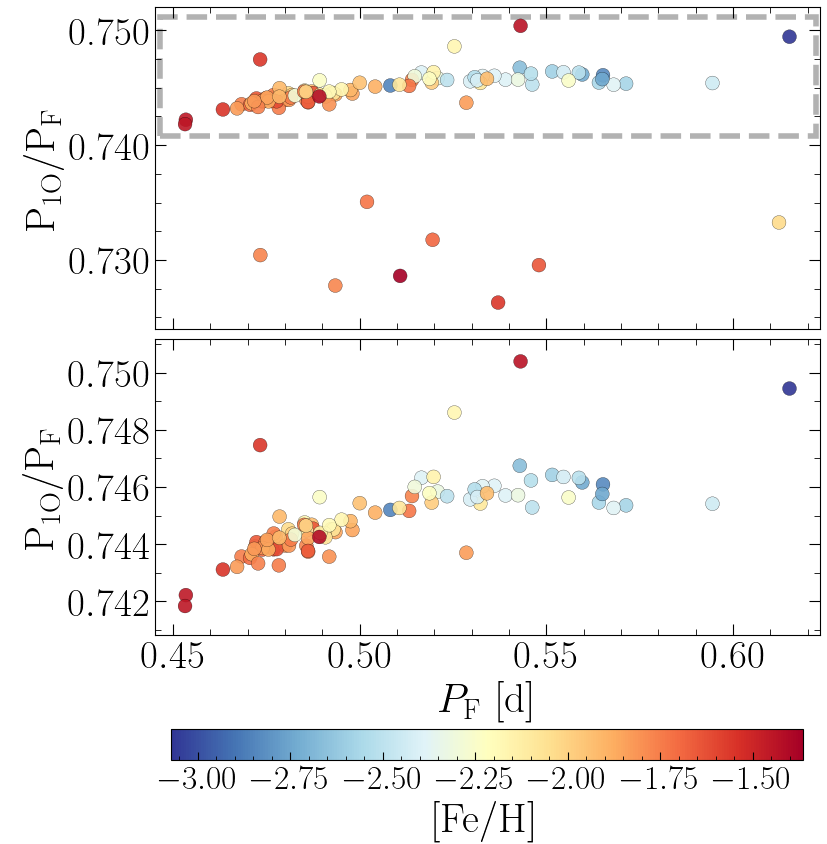}
\caption{
Period ratio (first-overtone period
to fundamental-mode period) versus fundamental period for the (field) RRd stars observed in DESI Y1, color-coded by [Fe/H].
The {\it top} panel depicts a zoommed out version of the diagram, showing the presence of RRd stars with non-canonical period ratios. 
The {\it bottom} panel highlights the region where the bulk of the distribution lies (shown with a dashed rectangle in the top panel), with period ratios between $\sim0.742$ and $\sim0.750$\,d. 
}
\label{fig:petersen}
\end{center}
\end{figure}

\begin{figure}
\begin{center}
\includegraphics[angle=0,scale=.44]{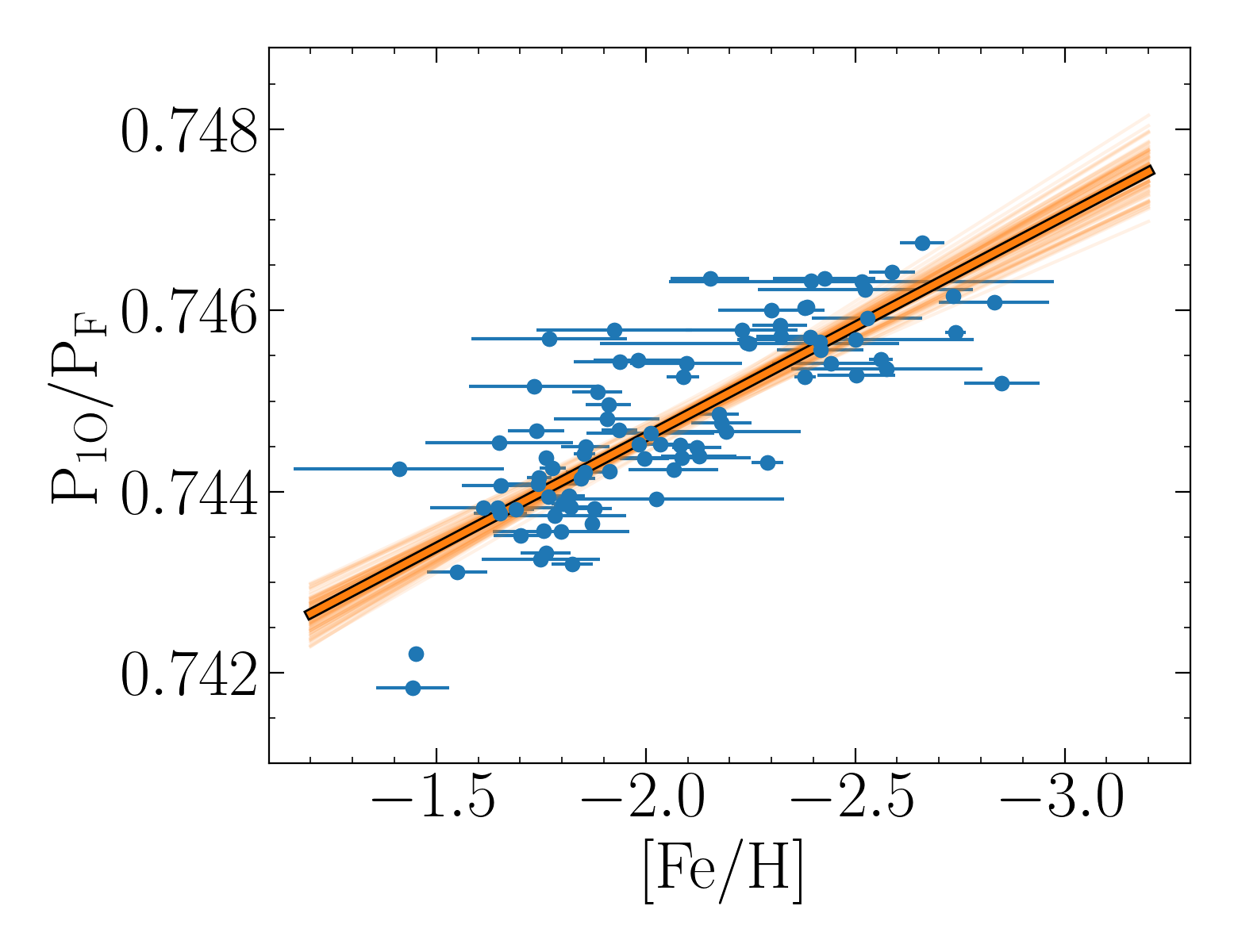}
\caption{
Period ratio as a function of [Fe/H] for the DESI Y1 RRd stars, fit with a linear function. 
In addition to the curve obtained from the median of the linear model parameters' posterior distributions, 100 curves drawn from the MCMC chains are shown to visualize the variation of the parameters. 
In this figure, the metallicity is shown decreasing from left to right for a direct comparison with Figure~9 in \citet{Braga2022}.
}
\label{fig:rrds}
\end{center}
\end{figure}

As pointed out by \citet{Soszynski2019}, the incidence rate of double-mode RRLs displays indications of a strong function of metallicity \citep[see e.g.,][]{Soszynski2014,Soszynski2016,Jurcsik2015,Varma2024}.
Moreover, the iron abundance of RRd stars plays a role in the position of these stars in the Petersen diagram, as suggested by previous studies \citep[e.g.,][]{Bono1996,Popielski2000} and recently  empirically characterized by  \citet{Braga2022}, \citet{Chen2023}, and \citet{Nemec2024}. 
Here, we take advantage of the sizable sample of field RRd stars observed by DESI in Y1 (107 stars) to further investigate this dependency.

Figure~\ref{fig:petersen} shows the Petersen diagram for the RRd stars in our sample. 
A metallicity gradient along the canonical sequence is clearly seen in this figure, with the more metal rich RRd stars at present at low $P_{\rm 1O}/P_{\rm F}$ ($\lesssim0.744$, or $P_{\rm F}<0.5$\,d), and the more metal poor stars displaying higher period ratios ($\gtrsim0.745$) with longer fundamental periods. 
We display the position of these stars in the Bailey diagram in Figure~\ref{fig:bailey}.
Additionally, the figure shows that at least seven anomalous RRd stars lie below the canonical RRd sequence (with $0.725<P_{\rm 1O}/P_{\rm F}<0.735$), and at least three lie above it (with $P_{\rm 1O}/P_{\rm F}>0.747$).
For all of the anomalous RRd stars, our measured [Fe/H] is $>-1.77$\,dex, 
and half of them exhibit GSE-like chemodynamics (with probabilities $p_{\rm GSE}>0.5$). 
In contrast, the fraction of classical RRd stars likely associated with GSE ($p_{\rm GSE}>0.5$) is only 18\%. 
We note that all of the observed [Fe/H] of our RRd sample, including those of anomalous RRd stars, lies within the expectations from 
theoretical models \citep[from stellar mass-metallicity-period ratio correlations; see e.g.,][]{Marconi2015}. 
It is important to mention that the [Fe/H] distribution of these anomalous RRd stars is significantly narrower than that of classical RRd stars ($-1.77 < {\rm [Fe/H]} < -1.36$\,dex vs. $-2.85 < {\rm [Fe/H] }< -1.41$\,dex, respectively).  
Using the well-defined mass-metallicity-period-ratio relation from \citet{Marconi2015}, we can constrain the mass of our sample of canonical RRd stars to $>0.69$\,M$_\odot$ (our sample extends to lower metallicities than those in their models), and the mass of the anomalous RRd subsample to the range 0.68-0.77\,M$_\odot$.

In Figure~\ref{fig:rrds}, we display the period ratios of our RRd stars as a function of [Fe/H]. 
The figure shows a relatively large scatter of  $P_{\rm 1O}/P_{\rm F}$ at a given [Fe/H], of $\sim0.0015$. 
We note that, as discussed by \citet{Popielski2000} and \citet{Marconi2015}, the scatter in period ratio at a given [Fe/H] could be explained by the spread in stellar mass of RRd stars, and that more massive RRd stars are expected to show larger period ratios at a given fundamental-mode period \citep[see e.g.,][]{Alcock1999,Kovacs1999,Soszynski2011}. 
Using the Python package \code{emcee}, we look for the best fitting parameters for the period ratio-[Fe/H] relation for classical RRd stars adopting a linear dependence and taking the uncertainties in [Fe/H] into account. 
We use \code{scipy}'s \code{optimize} module to find initial guesses for \code{emcee}, maximizing the likelihood of the parameters to be fitted 
and adopting the results from \citet{Braga2022} as initial guesses for the routine. 
Since the uncertainties in [Fe/H] are non-negligible, we adopt an orthogonal-distance likelihood, using the perpendicular distances of the data points from the fitted relation.
For \code{emcee}, we adopt uniform priors, namely $\mathcal{U}(0,2)$ and $\mathcal{U}(-0.5,0.5)$ for the zeropoint and the slope of the linear model, respectively.   
From this analysis, we obtain

\begin{equation}
\begin{array}{ll}
P_{\rm 1O}/P_{\rm F} & = 0.7397^{+0.0004}_{-0.0004} - 0.0024^{+0.0002}_{-0.0002}\ {\rm [Fe/H]}
\end{array}
\label{eq:perratio_feh_lin}
\end{equation}

From this equation, we note the remarkable agreement with the results from \citet{Braga2022}, $P_{\rm 1O}/P_{\rm F} = (0.7404 \pm 0.0007) - (0.0025 \pm 0.0004) \ {\rm [Fe/H]}$, 
particularly considering the different approaches used to derive [Fe/H] (in their case, based on the newest calibration of the $\Delta$S method; \citealt{Crestani2021a}).  
We note, however, that the quality of the fits can be largely affected by the presence or absence of RRd stars with measured [Fe/H] $<-2.7$ and $>-1.7$. 
Thus, larger samples of RRd stars in the metal-poor and metal-rich end of the [Fe/H] distribution are likely required to draw firmer conclusions on the validity of this model to describe the data.

\begin{figure}
\begin{center}
\includegraphics[angle=0,scale=.335]{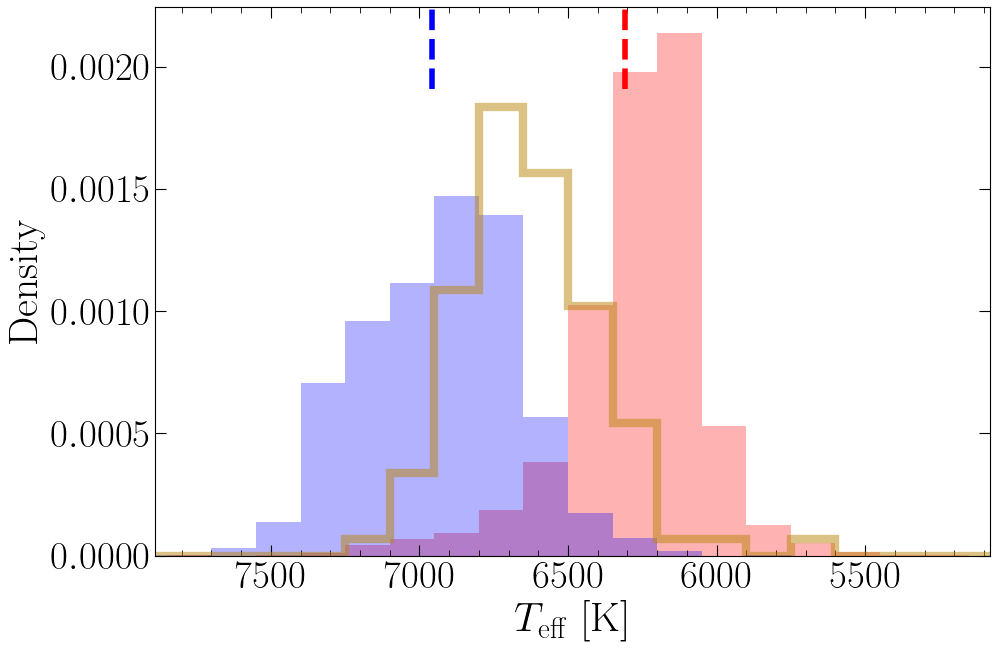}
\includegraphics[angle=0,scale=.335]{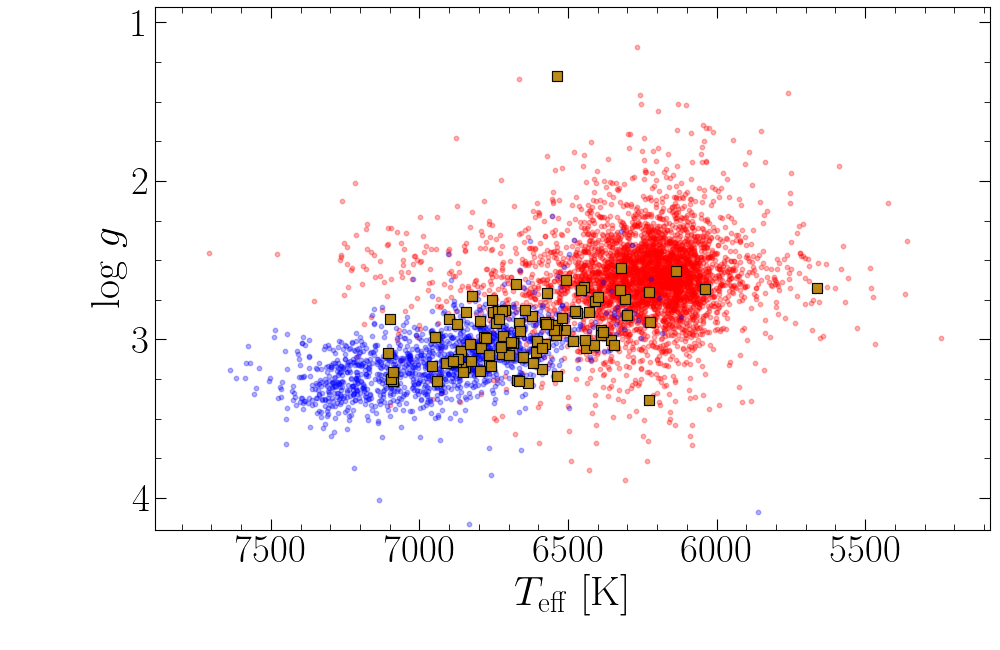}
\caption{
{\it Top}: Normalized distribution of (systemic) effective temperatures for the different RRL sub-types in our field sample, in \teff\ bins of $150$\,K. 
Vertical dashed lines depict the median of the 
for each sub-type. 
The distribution is largely concentrated between \teff\ of 5730 and 7520\,K, as indicated by the 1st and 99th percentiles of the RRab and RRc subsamples (respectively).
{\it Bottom}: Kiel diagram of the DESI field RRLs. 
This figure shows that the bulk of the RRab and RRc distributions occupy distinct regions of the \teff\ vs. log $g$ space, with the limit lying close to $\teff \sim 6500$\,K and log $g \sim 3$. 
In both panels, the bulk of the distribution of double mode pulsators is located between RRab and RRc stars in \teff. 
We note the presence of an outlier in the RRd distribution, resulting from an inaccurate model fit by the RVS pipeline  (its $\log g$ and \teff\ measured by DESI's SP pipeline are $3.2$ and $6700$\,K, respectively, consistent with the rest of the RRd distribution). 
}
\label{fig:teff_IS}
\end{center}
\end{figure}

\section{The edges of the RR Lyrae instability strip}
\label{sec:IS}

Since RRLs occur in a specific region of the HR diagram (where the horizontal branch crosses the instability strip), characterizing the morphology of their luminosity and \teff\ distribution in this region, both empirically and theoretically, plays an important role in our understanding of stellar evolution and the use of RRLs as population tracers. 
Indeed, the position of the blue and red edges of the RRL instability strip provide key insight on the theory of stellar pulsations, as specific physical conditions are required to drive pulsations in the stellar interiors through their heat mechanism, that is, the $\kappa$ and $\gamma$ mechanisms \citep[see e.g.,][]{Bono1994}.    
The blue edge's location is relevant since, if a star is too hot, a partial ionization zone of H or He (the responsible of the $\kappa$ and $\gamma$ mechanism-driven pulsations in RRLs) might not be present in the star, or might be located too far out in the star to trigger or maintain pulsations \citep{Iben1971,Catelan2015}.
Similarly, the position of the red edge is important to understand the limit in temperature in which stellar convection and its efficient transport of energy quenches the pulsations \citep[e.g.,][]{Schatzman1956,Xiong1982,Bono1995}. 
However, no large scale empirical study of the topology of the instability strip using RRLs (especially with homogeneously-derived spectroscopic properties) has been performed to date.

The rich and homogeneous DESI dataset offers an excellent opportunity to constrain empirically, for the first time using a large number of stars and pulsation-corrected \teff, the behavior of the hot and cold ends of the RRL instability strip, and their dependence on [Fe/H]. 
We note that it is expected that the colors (and temperatures) of RRLs during a considerable portion of their pulsation cycle (e.g., prior minimum light for RRab) lie outside the instability strip \citep[see e.g.,][]{For2011}.
Therefore, the observation phase of the stars must be considered when estimating the instability strip boundaries.
Here, we use the RRL effective temperatures derived in M25 to study these boundaries.
These estimates of the stars' mean \teff\ are computed mainly from single epoch spectra, accounting for cyclic variations by adopting \teff\ curve corrections. 
Additionally, we remove stars with \teff\ smaller than 4800\,K and hotter than 8250\,K, as we consider them outliers of the \teff\ distribution.

\begin{figure}
\begin{center}
\includegraphics[angle=0,scale=.335]{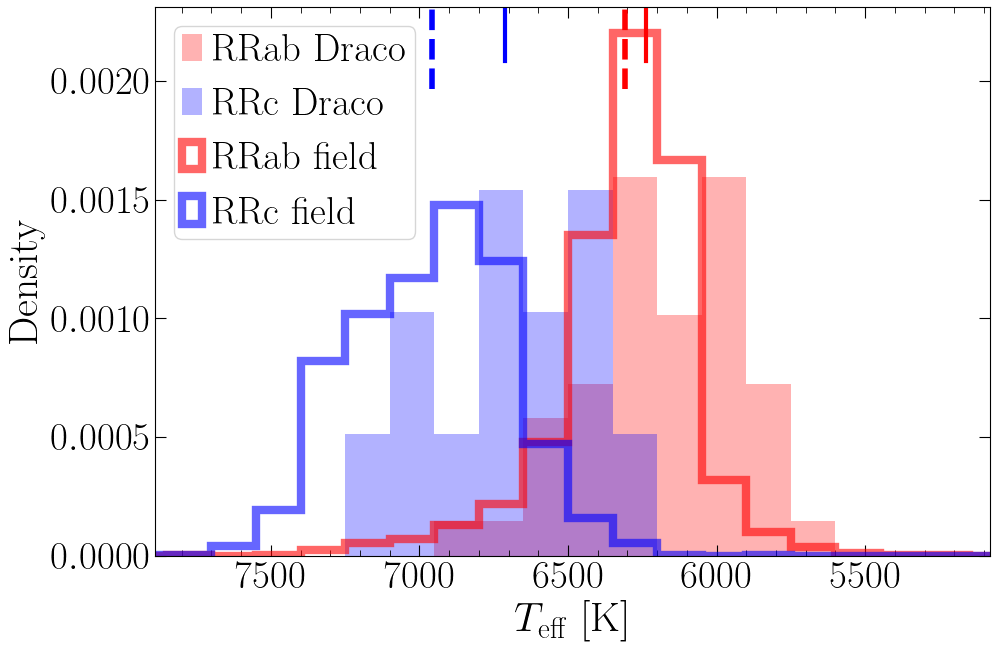}
\caption{
Same as Figure~\ref{fig:teff_IS} ({\it top}) but for the 59 RRls in Draco (filled histograms). 
Unfilled histograms are used to show the $T_{\rm eff}$ distribution of RRab and RRc in our field sample.
Solid vertical lines represent the median of the distributions for Draco's RRLs, whereas the values from the field sample are shown with dashed lines (as shown in Figure~\ref{fig:teff_IS}). 
The lower median \teff \ of the RRab and RRc in Draco  ($6187^{+248}_{-232}$ and $6735^{+290}_{-261}$\,K, respectively) as compared with those from the field sample ($6228^{+181}_{-182}$ and $6931^{+293}_{-245}$\,K) 
indicate the difference in the \teff \ distribution shape as a function of metallicity.
}
\label{fig:teff_IS_draco}
\end{center}
\end{figure}

Figure~\ref{fig:teff_IS} shows the distribution of mean effective temperatures for the 5,438 halo field RRLs in our sample with reliable temperature estimations, separated by RRL type. 
The median of the \teff\ distributions shown in the figure is $6224^{+181}_{-182}$\,K for RRab (where the errors denote the 84th and 16th percentiles of the distribution, respectively), $6931^{+293}_{-245}$\,K for RRc, and $6676^{+227}_{-224}$\,K for RRd variables. 
For a comparison with the \teff\ of RRLs in a more metal-poor population, we show in Figure~\ref{fig:teff_IS_draco} the \teff\ distribution of the 59 RRLs in Draco (${\rm [Fe/H]}\sim -2.0$\,dex), where the median \teff\ of the RRab and RRc is $6187^{+248}_{-232}$ and $6735^{+290}_{-261}$\,K, respectively.
Thus, in these plots we can see the effects of metallicity in the shape and position of the distributions.
From these distributions, we obtain the 1st percentile of the RRab \teff\ distribution and the 99th percentile of the RRc \teff\ distribution, 
as limits representing the temperature range out of which the number of stars becomes comparatively negligible. 
We find that the observed \teff\ distribution boundaries at the cold and hot ends are
$\sim5730$ and $7520$\,K, respectively. 
For the Draco-only sample, these values are 
$\sim5690$ and $7220$\,K. 
A more detailed inspection of the variation of the instability strip edges with metallicity is provided in Section~\ref{sec:ISvsFEH}, including Bayesian modeling, uncertainty deconvolution, and a comparison with previous results.

\subsection{The boundaries of the instability strip and [Fe/H]}
\label{sec:ISvsFEH}

Here, we quantify the dependence of the boundaries of the instability strip on metallicity,
with the goal of 
constraining its 
red edge (RE) and blue edge (BE)  
from our pulsation-corrected \teff.
To enable a more robust analysis, we only consider stars with uncertainties in \teff\ smaller than 300\,K (which resulted in 22 RRab stars and one RRc star removed from the full sample).
Furthermore, in order to reduce the effect of substructures in \teff\ vs. [Fe/H] space that could bias the statistics computed in the following subsections, we remove RRLs in globular clusters, dwarf galaxies, the Sgr stream, and GSE (for which we remove stars with membership probability $>0.50$ following the definition from M25).

\subsubsection{Trapezoid model}
\label{sec:trapezoid}

In our first approach, we model the intrinsic temperature distribution as a trapezoidal probability density function with edges varying linearly with [Fe/H].
This is done separately for RRab and for RRc stars, and we treat the cooler 
edge of the RRab distribution and the hotter 
edge of RRc stars as the instability strip edges. 
In what follows, we denote [Fe/H] as $y$ for simplicity in the notation.

In our parametrization, the 
cool and hot edges
of the \teff\ distribution of an RRL type vary linearly with $y$:

\textbf{\begin{equation}
\begin{array}{ll}
T_c(y) = x_{c0} + x_{c,y}\, y\\
T_h(y) = x_{h0} + x_{h,y}\, y\\
\end{array}
\label{eq:IS_trapezoid1}
\end{equation}}

\noindent where $x_{c0}$ and $x_{h0}$ are the zero point
of the cool and hot 
edges, $x_{c,y}$ and $x_{h,y}$ represent the slope of the edge with [Fe/H], and $T_{\mathrm{c}}(y)<T_{\mathrm{h}}(y)$.
Additionally, we adopt two positive softening lengths of the trapezoid $s_c$ and $s_h$, which determine the widths of the linear ramps at 
the cool and hot edges 
and must satisfy

\textbf{\begin{equation}
\begin{array}{ll}
s_c + s_h \le x_h(y) - x_c(y)
\end{array}
\label{eq:IS_trapezoid2}
\end{equation}}

\noindent Therefore, the model parameters are:

\begin{equation}
\begin{array}{ll}
\boldsymbol{\theta}
=
\left(
x_{c0},
x_{h0},
x_{c,y},
x_{h,y},
s_c,
s_h
\right)
\end{array}
\label{eq:IS_trapezoid3}
\end{equation}

For fixed $y$ the intrinsic probability density of \teff\ (denoted $T$ for simplicity in the equations below), $p_{\mathrm{int}}(T \mid y, \boldsymbol{\theta})$, is a linear trapezoid defined on the interval [$T_c(y)$,$T_h(y)$], and is defined by

\textbf{\begin{equation}
\begin{array}{ll}
p_{\mathrm{int}}(T \mid y, \boldsymbol{\theta}) =
\begin{cases}
0, & T < T_c(y), \\[6pt]
\displaystyle
\frac{h(y)}{s_c}
\big(T - T_c(y)\big),
& T_c(y) \le T < T_c(y) + s_c, \\[10pt]
h(y),
& T_c(y) + s_c \le T \le T_h(y) - s_h, \\[10pt]
\displaystyle
\frac{h(y)}{s_h}
\big(T_h(y) - T\big),
& T_h(y) - s_h < T \le T_h(y), \\[10pt]
0,
& T > T_h(y).
\end{cases}
\end{array}
\label{eq:IS_trapezoid4}
\end{equation}}

\noindent where $h(y)$ is a normalization term that ensures

\begin{figure}
\begin{center}
\includegraphics[angle=0,scale=.36]{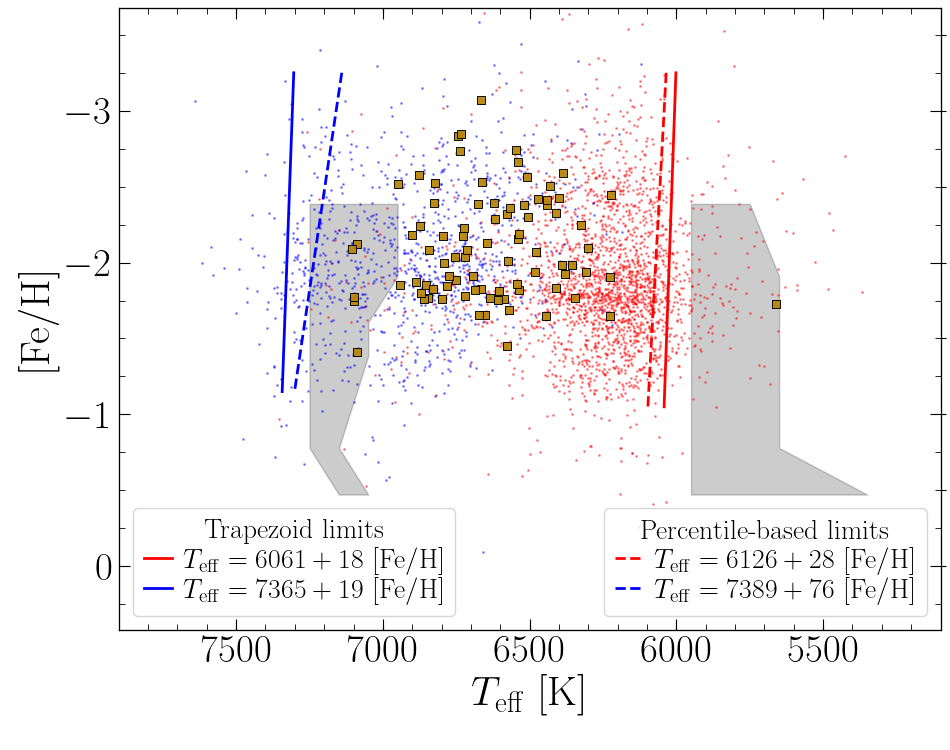}
\caption{
Similar to Figure~\ref{fig:perampmetcorrelation}, but depicting the mean \teff\ for the stars in our sample against their RVS iron abundances. 
In this diagram, RRc stars are located in the hotter side of the RRL \teff\ distribution (blue markers), whereas RRab (red markers) lie toward the cooler edge of the instability strip. 
Brown markers represent RRd stars, which lie predominantly in the region between RRab and RRc.  
Filled and dashed lines show  the results of our trapezoid model and of the linear fits to the 1st and 99th running percentiles of the deconvolved \teff distributions. 
The theoretical predictions for the edges of the RRL instability strip provided by \citet{Marconi2015} for different combinations of stellar masses, luminosities, and chemical abundances are shown as shaded regions. 
The leftmost borders of these regions represent the instability strip edges for the ZAHB and masses between 0.54-0.80\,M$_{\odot}$. 
}
\label{fig:teffmetcorrelation}
\end{center}
\end{figure}

\begin{equation}
\begin{array}{ll}
\int_{-\infty}^{\infty} p_{\mathrm{int}}(T \mid y,\boldsymbol{\theta}) \, dT = 1
\end{array}
\label{eq:IS_trapezoid5}
\end{equation}

We assume Gaussian measurement uncertainties in the observed values of \teff, 

\begin{equation}
\begin{array}{ll}
T_i^{\mathrm{obs}} \sim \mathcal{N} \left( T_i, \sigma_{T,i}^2 \right). 
\end{array}
\label{eq:IS_trapezoid6}
\end{equation}

\noindent Thus, the convolved observed density can be written as

\begin{equation}
\begin{array}{ll}
p_{\mathrm{obs}}(T_i^{\mathrm{obs}} \mid y_i, \sigma_{T,i}, \boldsymbol{\theta}) = \\ \ \ \ \ \ \ \ \ \ \ \ \ \ \ \ \ \ \int p_{\mathrm{int}}(T \mid y_i, \boldsymbol{\theta}) \, \mathcal{N}(T_i^{\mathrm{obs}} \mid T, \sigma_{T,i}^2) \, dT
\end{array}
\label{eq:IS_trapezoid7}
\end{equation}

\noindent and the likelihood of the model is defined as

\begin{equation}
\begin{array}{ll}
\mathcal{L}(\boldsymbol{\theta}) = \prod_{i=1}^{N} p_{\mathrm{obs}}(T_i^{\mathrm{obs}} \mid y_i, \sigma_{T,i}, \boldsymbol{\theta}).
\end{array}
\label{eq:IS_trapezoid8}
\end{equation}

To find the best fitting parameters of this model, we use \code{emcee} with \code{scipy}'s \code{optimize} module to find initial guesses. 
We adopt independent priors on all parameters within physically admissible ranges, including $s_c>0$, $s_h>0$. 
If we define $\Delta_T = \max \left( 300,\; x_{\max}-x_{\min} \right)$ from the minimum and maximum observed \teff used to fit the model, the priors are the following:

\begin{equation}
\begin{array}{ll}
x_{c0} \sim \mathcal{U}\!\left(x_{\min}-0.8\,\Delta_T,\; x_{\max}\right)\\
x_{h0} \sim \mathcal{U}\!\left(x_{\min},\; x_{\max}+0.8\,\Delta_T\right)\\
\ln s_c \sim \mathcal{U}\!\left(\ln 5,\; \ln(0.9\,\Delta_T)\right)\\
\ln s_h \sim \mathcal{U}\!\left(\ln 5,\; \ln(0.9\,\Delta_T)\right)\\
x_{c, y} \sim \mathcal{N}\!\left(0,\sigma_{\mathrm{slope}}^2\right)\\
x_{h, y} \sim \mathcal{N}\!\left(0,\sigma_{\mathrm{slope}}^2\right)
\end{array}
\label{eq:IS_trapezoid9}
\end{equation}

with

\begin{equation}
\begin{array}{ll}
\sigma_{\mathrm{slope}} = 0.18\,\frac{\Delta_T}{y_{\max}-y_{\min}}
\end{array}
\label{eq:IS_trapezoid10}
\end{equation}

\noindent where $y_{\min}$ and $y_{\max}$ are the minimum and maximum observed [Fe/H] in the sample.
We run \code{emcee} using 3000 iterations, 30 walkers, 1000 burn-in steps, and a thinning factor of 10, resulting in a final chain of effective samples.
The best fitting parameters are then obtained as the median of the posterior distributions, with uncertainties corresponding to their 16th and 84th percentiles. 
Our results, presented in Table~\ref{tab:params_IS_trapezoid}, and displayed in Figures~\ref{fig:teffmetcorrelation} and \ref{fig:cornerplots_trapezoid} (Appendix),
can be summarized in the shape of Equation~\ref{eq:IS_trapezoid1} as

\begin{equation}
\begin{array}{ll}
T_{\rm eff, trap}\ {\rm (RE)\ [K]} = 6061^{+29}_{-29} + 18^{+13}_{-13}\ {\rm [Fe/H]}\\
T_{\rm eff, trap}\ {\rm (BE)\ [K]}  = 7365^{+83}_{-62} + 19^{+25}_{-24}\ {\rm [Fe/H]},\\
\end{array}
\label{eq:IS_trapezoid11}
\end{equation}

\noindent where $T_{\rm eff, trap}\ {\rm (RE)}$ corresponds to $T_c(y)$ for RRab stars, and $T_{\rm eff, trap}\ {\rm (BE)}$ to $T_h(y)$ for RRc stars. 
If we instead define the boundaries to include half of the ramp widths, i.e., decreasing the limits by $0.5\ s_c$ and $0.5\ s_h$ (for the red and blue edges, respectively) we find

\begin{equation}
\begin{array}{ll}
T_{\rm eff,trap\ ramp}\ {\rm (RE)}\ [\rm K] = 6079^{+26}_{-26} + 18^{+13}_{-13}\ [\rm Fe/H] \\
T_{\rm eff,trap\ ramp}\ {\rm (BE)}\ [\rm K] = 7294^{+48}_{-54} + 19^{+25}_{-24}\ [\rm Fe/H].  
\end{array}
\label{eq:IS_trapezoid12}
\end{equation}

\noindent The corresponding trapezoid plateau boundaries ($T_c(y) + s_c$ and $T_h(y) - s_h$ for the red and blue edges, respectively) are: 

\begin{equation}
\begin{array}{ll}
T_{\rm eff,trap\ plat}\ {\rm (RE)}\ [{\rm K}] = 6097^{+29}_{-27} + 18^{+13}_{-13}\ [{\rm Fe/H}]\\
T_{\rm eff,trap\ plat}\ {\rm (BE)}\ [{\rm K}] = 7226^{+72}_{-112} + 19^{+25}_{-24}\ [{\rm Fe/H}].
\end{array}
\label{eq:IS_trapezoid13}
\end{equation}

\noindent In these equations, the zero points are computed from the MCMC chains and therefore reflect the posterior correlations between $x_{c,0}$ and $s_c$ (and $x_{h,0}$ and $s_h$; see Figure~\ref{fig:cornerplots_trapezoid}).

Figure~\ref{fig:trapezoid_realizations} (Appendix~\ref{sec:appendixA1}) shows 100 random draws from the MCMC chains evaluated at ${\rm [Fe/H]}=-1.5$ for RRab and RRc stars, illustrating the best-fit model and the posterior uncertainty in the inferred intrinsic distribution $p_{\mathrm{int}}(T_{\rm eff} \mid {\rm [Fe/H]},\,\boldsymbol{\theta})$. 
We note that a caveat of our approach is the assumed trapezoidal shape of the intrinsic temperature distribution. 
The true form of the distribution, and how sharp the physical edge is, is not known a priori, so the adopted parameterization is necessarily idealized. 
In stellar pulsation theory, the strip is defined by the transition between stable and unstable models in linear non-adiabatic calculations, while modern RRL instability strip computations incorporate nonlinear hydrodynamics and time-dependent convection \citep[see e.g.,][]{Catelan2015,Marconi2015,Netzel2024,Cruz-Reyes2024}.
Our adopted parameterization should therefore be regarded as a phenomenological description of the data rather than an  exact representation of the instability strip.

While the red (low-temperature) and blue (high-temperature) edges are relatively well constrained in position, their corresponding softening lengths are not equally well determined (see Table~\ref{tab:params_IS_trapezoid}). 
In both RRab and RRc samples, the posterior maximum for $s_c$ lies at the lower prior boundary, producing a truncated posterior toward small values and indicating weak constraints on the ramp structure. 
This is also the case for the softening length of RRc stars. 
For this reason, we also report the 95th percentile of these parameters as upper limits in Table~\ref{tab:params_IS_trapezoid}.  
Moreover, we find that for both RRab and RRc stars, $s_h$ exhibits comparatively large values, 
implying an extended high-temperature tail in the inferred intrinsic distributions. 
This likely reflects the substantial scatter of stars near the blue side of the strip and suggests that the model absorbs dispersion through an inflated ramp width. 

Taken together, these results indicate that the detailed shape of the strip boundaries remains only loosely constrained by the present data. 
As a result, the total width of the instability strip cannot be determined with high confidence within this model. 
The most robust information provided by the data is instead contained in the percentiles of the deconvolved intrinsic distribution.
In the next section, we focus on a percentile-based characterization, which is more directly data-driven and less dependent on the assumed functional form.

\begin{table*}\small
\caption{
Results of our trapezoidal model of the edges of the instability strip. 
For each of our model's parameters, we report the median of the posterior distributions, with uncertainties representing the 16th and 84th percentiles.
For parameters that are weakly constrained by our model, 
we also report the 95th percentile of their posteriors' distributions as upper limits. 
}
\label{tab:params_IS_trapezoid}
\begin{center}
\begin{tabular}{|c|c|c|c|c|c|c|}
\hline
\multicolumn{1}{|c|}{Type} &
\multicolumn{1}{c|}{$x_{c,0}$} &
\multicolumn{1}{c|}{$x_{h,0}$} &
\multicolumn{1}{c|}{$x_{c,y}$} &
\multicolumn{1}{c|}{$x_{h,y}$} &
\multicolumn{1}{c|}{$s_c$} &
\multicolumn{1}{c|}{$s_{h}$} \\
 & (K) &   (K) &  (K\, dex$^{-1}$) & (K\, dex$^{-1}$) &  (K) & (K)     \\     
\hline
  ab & $6061^{+29}_{-29}$ & $6751^{+39}_{-38}$ &  $18^{+13}_{-13}$ & $12^{+17}_{-18}$ & $32^{+30}_{-19}$ ($<85$) &  $604^{+34}_{-38}$ \\
   c & $6952^{+54}_{-56}$ & $7365^{+83}_{-62}$ & $203^{+26}_{-24}$ & $19^{+25}_{-24}$ & $55^{+54}_{-36}$ ($<151$) & $122^{+185}_{-84}$ ($<444$) \\
\hline
\end{tabular}
\end{center}
\end{table*}

\subsubsection{Deconvolution of the observed \teff\ distribution}
\label{sec:deconvolved}

In addition to the Bayesian model used in Section~\ref{sec:trapezoid}, we follow a data-driven approach to infer the intrinsic \teff\ distribution of our sample.
Similar to the methodology described in Section~\ref{sec:OoAndBailey}, this time using pulsation-corrected \teff\ instead of periods and amplitudes, we split our sample of RRLs in [Fe/H] bins containing the same number of stars (in the [Fe/H] range [$-2.79$,$-1.19$]\,dex) and compute statistics in each subsample.

We adopt a non-parametric deconvolution approach based on the Richardson–Lucy (RL) iterative algorithm \citep{Richardson1972,Lucy1974}, applied separately to RRab and RRc stars. 
In each metallicity bin, the observed \teff\ distribution is represented by a normalized histogram and modeled as the convolution of an intrinsic probability distribution function (PDF) with measurement uncertainties. 
Because the temperature errors are heteroscedastic, we approximate them within each bin by a single effective Gaussian kernel whose width is taken as the median \teff\ uncertainty. 
The RL iteration is initialized with the observed histogram  and computed using Fast Fourier Transform-based convolutions implementation in {\tt scipy}. 
The solution is evolved for a fixed number of iterations $N_{\rm iter}$, which acts as a regularization parameter limiting noise amplification.

To assess the adequacy of the recovered intrinsic PDF, we perform forward modeling by drawing Monte Carlo samples from the RL solution, reconvolving them with the empirical distribution of temperature uncertainties in the bin. 
The resulting distribution is compared to the observed \teff\ histogram to verify consistency, which is achieved with 30 and 10 iterations for RRab and RRc stars. 
From the intrinsic PDF, we compute the 1st, 5th, 16th, 50th, 84th, 95th, and 99th percentiles, 
and adopt the 1st percentile of the intrinsic RRab \teff\ distribution and 99th percentile of the intrinsic RRc \teff\ distribution as operational, percentile-based edges of the instability strip  (per metallicity bin). 
We acknowledge that this choice of percentile thresholds is inherently arbitrary, and the obtained instability strip edges should be regarded as a pragmatic, data-driven definition rather than a physically sharp boundary.

Similar to the methodology used in Section~\ref{sec:OoAndBailey}, 
we find linear trends in the reported 
percentile-based \teff\ 
limits in the [Fe/H] range explored, hence we fit linear relations for them.  
The results of this exercise are shown in Figure~\ref{fig:teffmetcorrelation} and can be expressed as 

\begin{equation}
\begin{array}{ll}
T_{\rm eff}\ {\rm (RE)\  [K]}& = 6126\ (\pm 17) + 28\ (\pm 8)\ {\rm [Fe/H]},\\
T_{\rm eff}\ {\rm (BE)\  [K]}& = 7389\ (\pm 76) + 76\ (\pm 37)\ {\rm [Fe/H]}.
\end{array}
\label{eq:TEFFfeh_relation_3sig}
\end{equation}

\begin{table*}\small
\caption{
Statistics of the \teff\ distribution of RRab and RRc in ten metallicity bins, each containing the same number of stars ($N_{\rm bin}$). 
The columns q1, q5, q16, q50, q84, q95, and q99 correspond to the 1st, 5th, 16th, 50th, 84th, 95th, and 99th percentiles of the \teff\ distributions, respectively. 
}
\label{tab:params_IS}
\begin{center}
\begin{tabular}{|c|c|c|c|c|c|c|c|c|}
\hline
\multicolumn{1}{|c|}{Type} &
\multicolumn{1}{c|}{[Fe/H] range} &
\multicolumn{1}{c|}{q1} &
\multicolumn{1}{c|}{q5} &
\multicolumn{1}{c|}{q16} &
\multicolumn{1}{c|}{q50} &
\multicolumn{1}{c|}{q84} &
\multicolumn{1}{c|}{q95} &
\multicolumn{1}{c|}{q99} \\
     
 &   (dex) &   (K) &  (K) & (K) &  (K) & (K) &  (K)  &  (K)    \\     

\hline
     RRab & [$-2.49$, $-2.23$) & 6046 & 6091 & 6147 & 6220 & 6295 & 6356 & 6415 \\
($N_{\rm bin}=237$) & [$-2.23$, $-2.04$) & 6050 & 6117 & 6172 & 6255 & 6339 & 6400 & 6456 \\
          & [$-2.04$, $-1.91$) & 6080 & 6123 & 6162 & 6253 & 6324 & 6361 & 6420 \\
          & [$-1.91$, $-1.82$) & 6056 & 6102 & 6150 & 6243 & 6336 & 6395 & 6453 \\
          & [$-1.82$, $-1.75$) & 6070 & 6116 & 6179 & 6249 & 6330 & 6383 & 6425 \\
          & [$-1.75$, $-1.67$) & 6081 & 6132 & 6169 & 6244 & 6315 & 6375 & 6429 \\
          & [$-1.67$, $-1.54$) & 6075 & 6124 & 6164 & 6237 & 6313 & 6361 & 6453 \\
          & [$-1.54$, $-1.36$) & 6081 & 6127 & 6165 & 6230 & 6301 & 6357 & 6406 \\
      RRc & [$-2.65$, $-2.45$) & 6509 & 6573 & 6635 & 6784 & 6879 & 6977 & 7124 \\
($N_{\rm bin}=101$) & [$-2.45$, $-2.26$) & 6501 & 6593 & 6653 & 6808 & 7024 & 7101 & 7222 \\
          & [$-2.26$, $-2.10$) & 6607 & 6694 & 6777 & 6954 & 7088 & 7162 & 7279 \\
          & [$-2.10$, $-1.99$) & 6614 & 6684 & 6746 & 6893 & 7038 & 7156 & 7228 \\
          & [$-1.99$, $-1.91$) & 6618 & 6679 & 6751 & 6884 & 7041 & 7167 & 7261 \\
          & [$-1.91$, $-1.80$) & 6667 & 6707 & 6771 & 6867 & 7092 & 7194 & 7264 \\
          & [$-1.80$, $-1.69$) & 6649 & 6749 & 6809 & 6938 & 7084 & 7177 & 7280 \\
          & [$-1.69$, $-1.42$) & 6648 & 6704 & 6751 & 6882 & 7015 & 7122 & 7216 \\
\hline
\end{tabular}
\end{center}
\end{table*}

These equations show that the data-driven blue boundary of the instability strip is correlated with metallicity, while for the RE the correlation is negligible. 
Based on these definitions of the instability strip boundaries, we would expect the RE to lie 
at around 6080\,K 
for field stars in the inner halo ([Fe/H] $\sim-1.6$\,dex) and for a more metal-poor stellar population ([Fe/H] $\sim-2.0$\,dex). 
We note that, in the [Fe/H] range [$-2.0$,$-1.0$], the shape of the correlation found with this method for the RE is in remarkable agreement with that from our trapezoid model, with an offset of $\sim60$\,K.
On the hotter side, 
we find 
a BE close to 7270\,K
for [Fe/H] $\sim-1.6$\,dex and 
6240\,K 
for [Fe/H] $\sim-2.0$\,dex.

\subsubsection{Comparison with the literature}

\citet{Marconi2015} predicted the position of the edges of the instability strip using nonlinear convective hydrodynamical models of RRLs in a broad range in metal abundances ($Z=0.0001$--$0.02$). 
Figure~\ref{fig:teffmetcorrelation} also shows the region in [Fe/H] vs. \teff\ space covered by the discrete predicted values  of the first overtone blue edge and the fundamental red edge of the instability strip, as provided by \citet{Marconi2015} (Table~2 in their work).
These values include the instability strip edges at the zero-age horizontal branch (ZAHB) for stellar masses of 0.54-0.80\,M$_\odot$ (the leftmost limit of each shaded region in Figure~\ref{fig:teffmetcorrelation}; sequence A in their table), and those for a luminosity 0.1\,dex brighter than the ZAHB (sequence B), for the luminosity of central He exhaustion (sequence D), and for a 10\% less massive and 0.2\,dex brighter than the ZAHB model (sequence C).  

We find a good agreement between the predicted shape of the red edge at the ZAHB (i.e., their sequence A) and the shape of our derived 
limits in the entire [Fe/H] range of comparison. 
For the blue edge, the predicted boundary at the ZAHB lies closer to our percentile-based 
region rather than our broader trapezoid model (Equation~\ref{eq:IS_trapezoid11}).
The percentile-based limits are, however, consistent with the trapezoid model in the region between the half-ramp midpoint and the plateau boundaries (Equations~\ref{eq:IS_trapezoid12} and ~\ref{eq:IS_trapezoid13}), suggesting that both approaches delineate the same physical region of the instability strip.
We note that our measurements between [Fe/H] of $-2$ and $-1$ 
are in rough agreement with the results from \citet{Marconi2015}, which predict a roughly constant width (of the order of $\sim$1250-1400\,K; we find $\sim$1150-1300\,K for our trapezoid model and $\sim$1200-1350\,K for our percentile-based estimate) among similar chemical compositions. 
In Figure~\ref{fig:teffmetcorrelation}, we find that the instability strip as a whole moves towards cooler \teff\ with more metal-poor RRLs, although this trend is not statistically significant at the red end. 
This tendency is also observable if we consider non-corrected \teff\ from DESI's RVS or SP pipelines. 
Interestingly, a similar trend 
was identified by \citet{Luo2021} using $\sim400$ RRab in LAMOST DR6, and previously reported by \citet{Sandage1993}. 
We note that these results contrast with the predictions of \citet{Marconi2015} (available down to ${\rm [Fe/H]}\sim-2.20$\,dex) of a shift toward cooler \teff\ with increasing metal content. 
Potential explanations for this shift include a systematic lost of precision and accuracy for [Fe/H] and/or \teff\ in the very metal-poor regime ([Fe/H] $<-2$\,dex) in our data,
limitations and biases in the determination of pulsation-corrected \teff, 
or the unsuitability of the hydrodynamical models considered in this metallicity regime. 
The discrepancy could also be explained by the relative numbers of post-ZAHB stars at different metallicities 
or by the  lack of sizable samples of RRLs near the red edge of the fundamental-mode instability strip (which could cause the 1st percentiles to move towards hotter temperatures) 
and RRc near the blue edge of the instability strip (idem, but for the 99th percentiles).

In Table~\ref{tab:params_IS}, we report percentiles of 
of our deconvolved \teff\ distributions 
in different [Fe/H] bins (each containing the same number of stars).
The trend described above for the boundaries of the instability strip, which is more pronounced for RRc stars, can be deduced from the numbers shown in the table. 
These numbers do not show a consistent variation of the width of the RRab star \teff\ distribution with [Fe/H], and that is also the case for RRc stars.  
Moreover, we can infer details about the region that separates the RRab and RRc distributions in \teff. 
For instance, the 84th percentile \teff\ limit of RRab stars across [Fe/H] is most similar to the 16th percentile of RRc stars. 
In practice, this means that in terms of relative numbers, the region occupied by the hotter end of the RRab distribution contains a high fraction of RRc stars. 
Figure~\ref{fig:teffmetcorrelation} also shows that the bulk of the RRd \teff\ distribution lies in this region, where the fundamental and the first-overtone modes approach pulsationally stable nonlinear cycles \citep[the so-called ``OR'' region; see e.g.,][]{Marconi2015}. 

A number of previous studies have empirically estimated the effective temperature at the red-horizontal branch (RHB) boundary where the onset of convection suppresses stellar pulsations \citep[see e.g.,][]{Preston2006,For2010,For2011,Hansen2011}. 
\citet{Preston2006}, for instance, analyzed the temperature distribution of  24 field metal-poor ([Fe/H] $<-2$\,dex) red horizontal-branch stars and six of these stars in the very metal-poor globular cluster M15, combined with globular cluster RRL data, to infer the position of the fundamental red edge of the metal-poor RRL instability strip at $\sim6310\pm150$\,K. 
Later, \citet{For2011} used 2300 high-resolution spectra of 11 field RRab stars with $-2.0<$[Fe/H]$<-1.0$ to estimate the position of the instability strip red-edge to lie at $6280\pm30$\,K, similar to the results of \citet{Preston2006}, although a more metal-rich sample was used in the latter case. 
Based on large number statistics and our arbitrary definition of the instability strip edges, we estimate a RHB boundary in the range [6020, 6080]\,K for halo field stars. 
The origin of this discrepancy warrants further investigation with larger RRL samples with known \teff, improved \teff\ corrections and uncertainties, and tighter constraints on the evolutionary state of individual stars.

Lastly, we note in passing that \citet{Gillet2013} recognized that RRab exhibiting Blazhko modulation are preferentially located near the blue-edge of the RRL fundamental-mode instability strip (as opposed to non-Blazhko RRab often found closer to the red edge of the instability strip). 
We identify the investigation of the location of Blazhko vs non-Blazhko RRab in the HR diagram as a direct application of the DESI Y1 RRL catalog, which would also require a thorough photometric study of the complete sample.
We note, however, that a detailed analysis of this specific science case is not possible with {\it Gaia} data alone and is beyond the scope of this paper.

\section{Metal-rich RRL candidates}
\label{sec:youngAndMrich}

The traditional formation scenario for RRLs 
involves single-star evolution, where the fraction of horizontal branch stars that cross the instability strip (therefore, their overall pulsation properties) depend on their metallicity, initial mass, and wind mass-loss during the red giant phase.
This single evolution channel produces RRLs that are old ($>10$\,Gyr) and metal-poor ([Fe/H] $<-1$\,dex) Population II stars, which are those predominantly observed in the Galaxy. 
An additional formation channel involving binary system evolution was recently discussed by \citet{Bobrick2024} based on stellar evolution simulations from the Modules for Experiments in Stellar Astrophysics (MESA) code \citep{Paxton2011,Paxton2019}. 
In this scenario, a horizontal branch star that falls in the instability strip can be produced when one of the stars in the system strips material from its red giant branch companion (the RRL progenitor, which loses most of its mass during its red giant phase). 
The likelihood of this happening depends on the amount of stripped material, among other factors.
By performing a stellar population synthesis study, \citet{Bobrick2024} found that $\sim10$\% of the Galactic RRLs and the majority of those with observed high metallicities ([Fe/H] $\geq-1.0$, e.g., in the Solar neighborhood) can be formed in binary systems. 
Moreover, these authors predicted that all metal-rich RRLs have an A, F, G, or K-type companion with an orbital period between 900--1,900\,d, and that approximately one of every 1,000 RRLs can be a product of binary formation. 
We note, however, that to the best of our knowledge no strong empirical evidence has been found to support this formation scenario \citep[see e.g.,][]{Abdollahi2025}.

Metal-rich RRLs that were not formed via the canonical single evolution channel challenge our understanding of stellar evolution, binarity, and the formation of our Galaxy.  
These stars have been observed in the Galaxy with chemo-dynamical properties (iron abundances, spatial distribution, and kinematics) consistent with those of thin disk stars \citep[see e.g.,][]{Liu2013,Marsakov2019,Prudil2020,Zinn2020,Crestani2021b,Iorio2021}.
The (theoretical) scenario that describes the pulsational and evolutionary properties of metal-rich RRLs requires their progenitors to undergo substantial mass loss during their ascent to the red giant branch \citep[][]{Bono1997a,Bono1997b}, induced, for instance, by shock wave winds, radiation pressure, and/or high rotation rates \citep[see e.g.,][]{DCruz1996}. 
The binary system formation scenario provides a means for this considerable mass loss  
and also allows RRLs to be formed at young ages (anywhere between 1-10\,Gyr old), challenging the traditional view of RRLs as exclusively old stellar populations. 
Indeed, the age-metallicity relation obtained by \citet{Bobrick2024} for binaries in their MESA simulation (see their Figure 6) shows that, with high confidence, the metal-rich tail of the RRL distribution ([Fe/H] $>-0.5$) traces young populations (younger than 5\,Gyr). 
This is supported by the recent findings of \citet{Zhang2025}, who reported that RRLs with ${\rm [Fe/H]} > -0.5$\,dex display kinematics consistent with a population significantly younger than typically assumed for RRLs. 
Here, we discuss potential metal-rich candidates in the DESI Y1 RRL sample.

\begin{figure*}
\begin{center}
\includegraphics[angle=0,scale=.30]{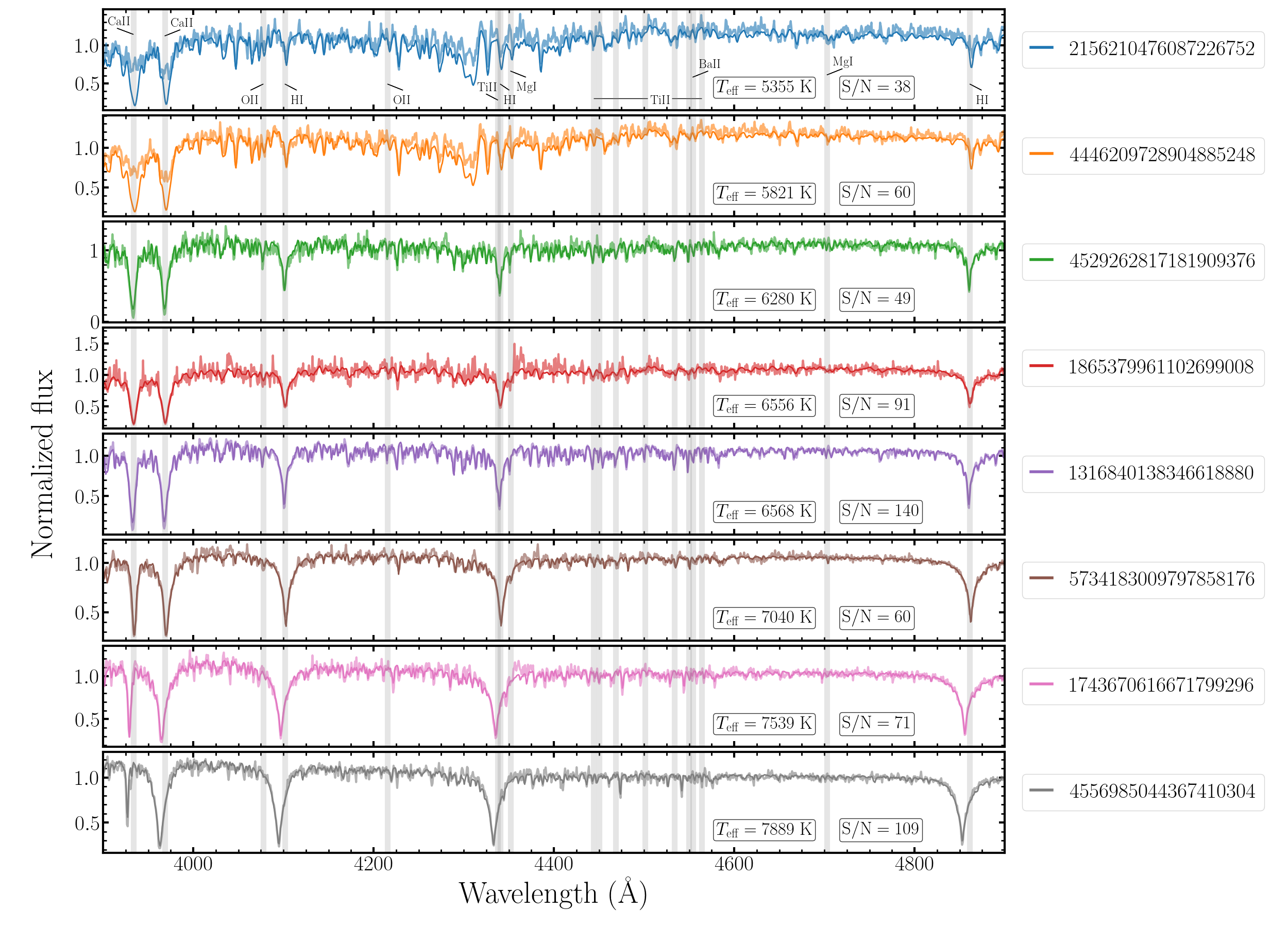}
\caption{
Spectra of the eight RRLs in the DESI catalog that display [Fe/H] $>-0.5$\,dex (from the RVS pipeline), which are described in Section~\ref{sec:youngAndMrich}. 
The observed spectra are depicted with thinner and more transparent lines, whereas the best-fitting RVSpec models are shown with thicker lines. 
The region displayed contains three absorption lines in the Balmer series (H$_\beta$, H$_\gamma$, H$_\delta$), in addition to calcium (Ca II H and Ca II K) and lines used by \citet{Medina2023}.
The signal-to-noise ratio of the spectra in the blue arm of the detector is shown as a reference. 
}
\label{fig:spectraMRYR}
\end{center}
\end{figure*}

\begin{figure*}
\begin{center}
\includegraphics[angle=0,scale=.32]{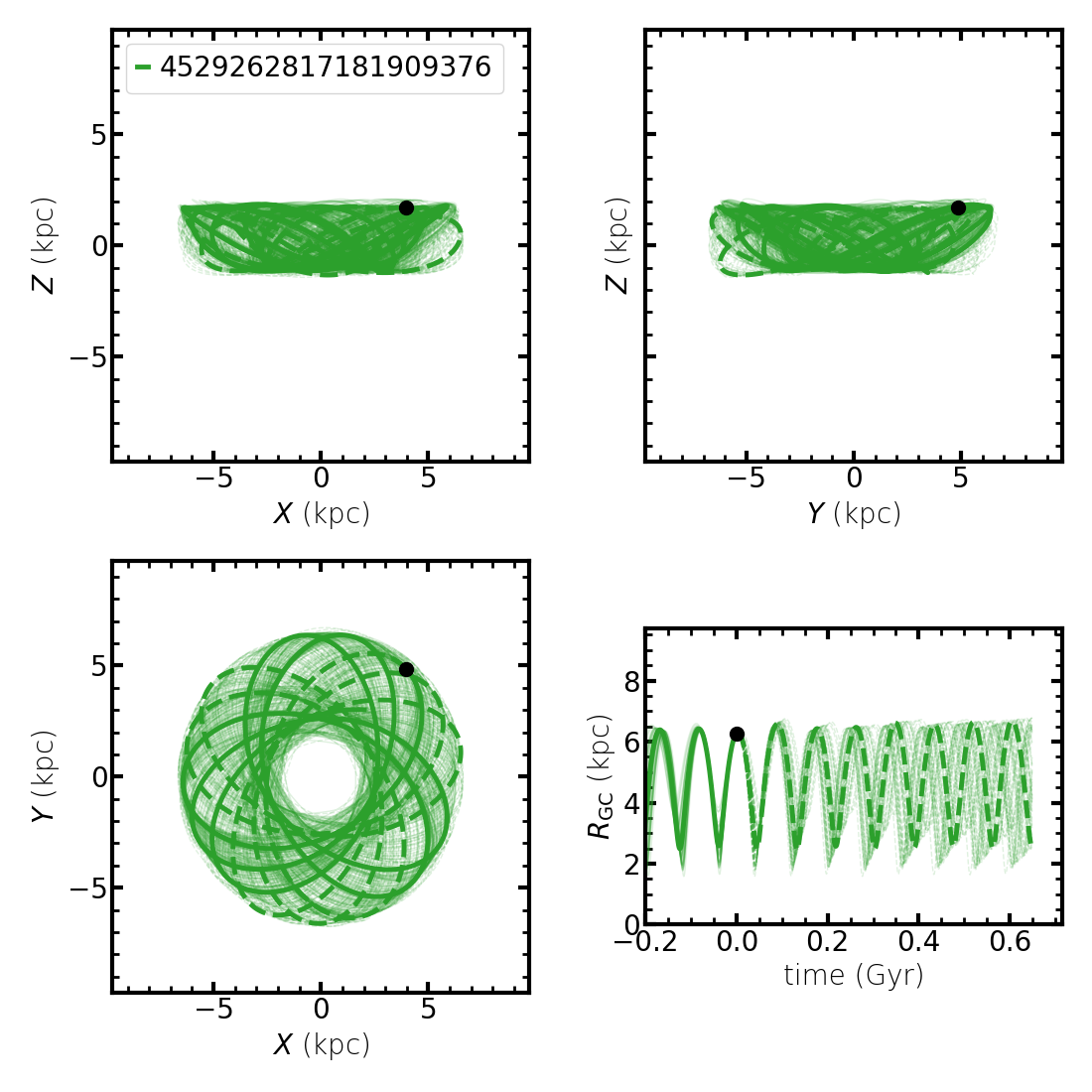}
\includegraphics[angle=0,scale=.32]{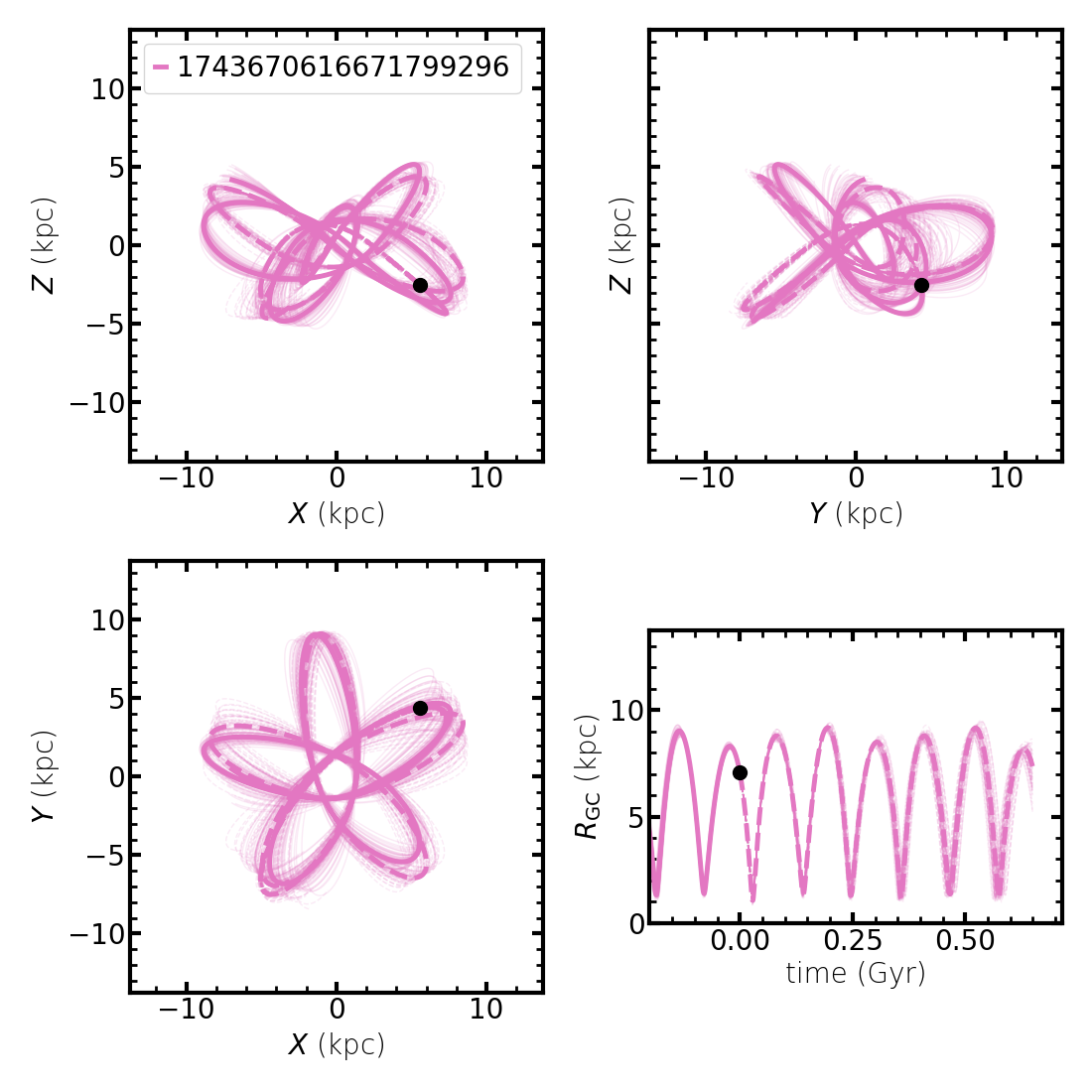}\\
\includegraphics[angle=0,scale=.32]{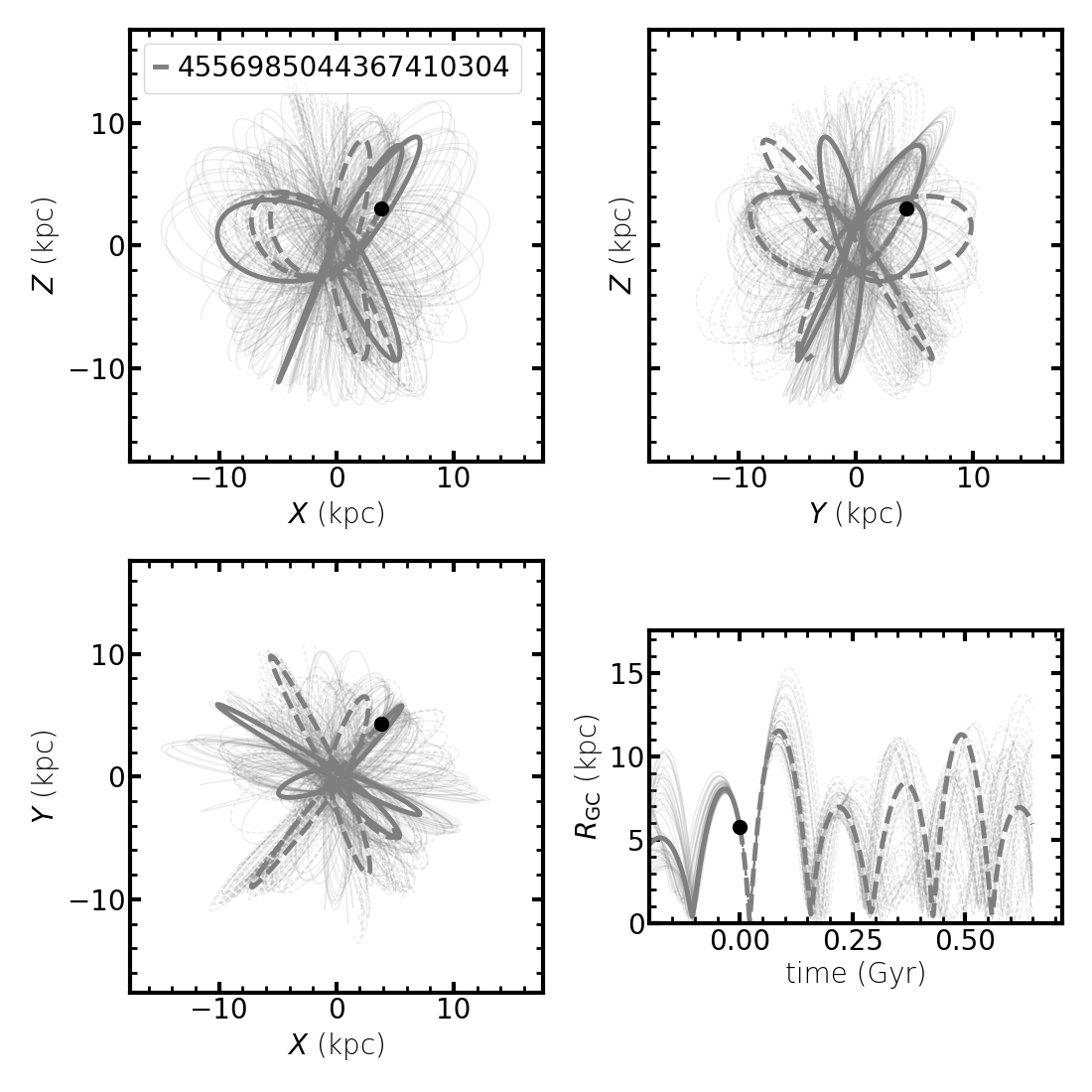}
\includegraphics[angle=0,scale=.32]{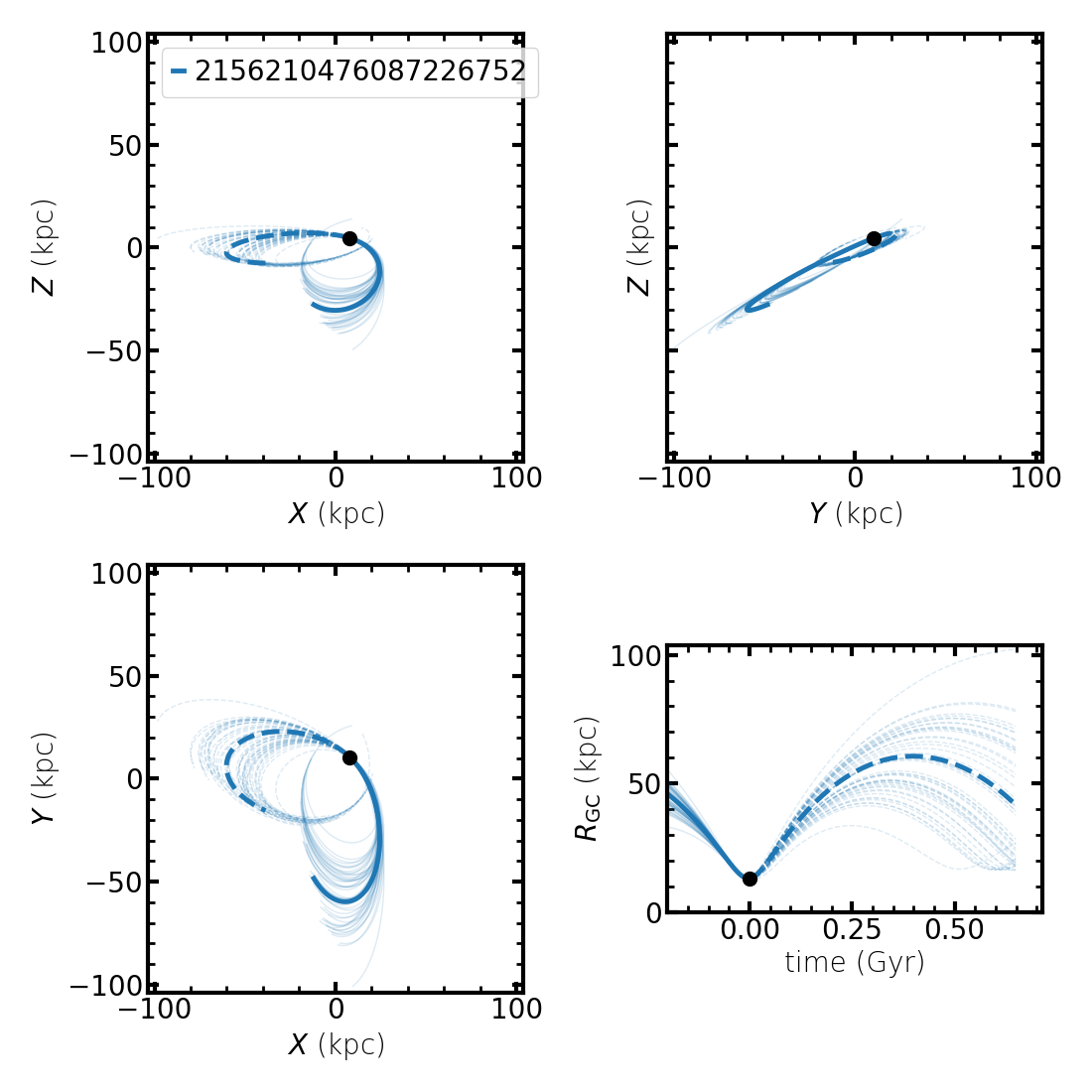}
\caption{
Orbits of four of the metal-rich candidates discussed in Section~\ref{sec:youngAndMrich}, in Galactocentric Cartesian coordinates ($X$, $Y$, and $Z$) and Galactocentric radius ($R_{\rm GC}$), integrated for 0.65\,Gyr.
In each panel, a solid line is used to represent the stars' orbits integrated backwards in time, and their inferred future trajectories are shown with dashed lines. 
These panels also depict a fraction of the orbits computed for the determination of the stars' orbital parameters (10\%) with transparent lines.
The two sets of panels on top are two examples of stars with disk-like kinematics, corresponding to the stars 4529262817181909376 and 1743670616671799296. 
We also show the orbit of a metal-rich candidates with signs of halo kinematics (4556985044367410304 and 2156210476087226752).
}
\label{fig:orbitsMRYR}
\end{center}
\end{figure*}

\subsection{Selection and kinematics of metal-rich candidates}

\citet{Bobrick2024} showed that regions containing metal-rich and young stars in dynamically cold orbits, like the thin disk, are good environments for the identification of  metal-rich and young RRLs.  
Thus, identifying metal-rich candidates in a survey that does not focus on these regions, like DESI, might provide useful constraints on the nature of binary star evolution, and Galactic dynamics. 
We employ the observed properties of DESI's metal-rich RRLs (systemic velocities, heliocentric distances, and proper motions) to investigate the nature of their kinematics and trace their origins back in time.

We consider stars with  ${\rm [Fe/H]}\geq-0.5$\,dex in the DESI RRL sample as candidate metal rich RRL. 
According to this criterion, nine stars fall into the metal-rich candidate category when using RVS-based iron abundances (${\rm [Fe/H]}_{\rm RVS}$), and a total of 63 RRLs exhibit ${\rm [Fe/H]}>-0.5$\,dex from either of the three methods used to determine metallicity for DESI's RRLs. 
Of the nine RRLs with ${\rm [Fe/H]}_{\rm RVS}>-0.5$\,dex, one star (with {\it Gaia} {\sc source\_id} 1209496845350566656) displays almost no clear absorption lines in its spectrum (its S/N is $<3$ at wavelengths $<5500$\,K), with the exception of the calcium triplet region (between 8498 and 8662\,\AA), where ${\rm S/N}\sim25$. 
A tenth star, classified in  {\it Gaia} as an RRc star (1736044064723548160) with low {\sc best\_class\_score} (previously discussed as a contaminant in our SASP sample; see Section~\ref{sec:haspsasp}), also displays [Fe/H]$_{\rm RVS}>-0.5$\,dex. 
For this star, we find large proper motions with small uncertainties (with proper motions of $-12.39\pm0.79$ and $-21.24\pm0.83$\,mas\,yr$^{-1}$ in right ascension and declination) for its distance ($d_{\rm H}=54.4\pm1.0$\,kpc), in addition to a high eccentricity ($\sim0.99$) and negative parallax ($-2.60\pm0.9$\,mas\,yr$^{-1}$).  
Hence, this star is most likely a contaminant in the sample (i.e., not an RRL). 
The spectra of the eight remaining stars, as well as their best-fitting models (from the RVS pipeline), are displayed in Figure~\ref{fig:spectraMRYR}, and their main properties are reported in Table~\ref{tab:params_MRYR}.

Of the eight stars with good spectra shown in Figure~\ref{fig:spectraMRYR}, 
two show [Fe/H] metallicities from DESI's SP pipeline also consistent with the $>-0.5$\,dex threshold (4556985044367410304 and 4529262817181909376), and two show $\Delta$S-method metallicities larger than $-0.5$\,dex (4529262817181909376 and 1865379961102699008). 
Of these, the star 4529262817181909376 shows ${\rm [Fe/H]}>-0.5$\,dex from all three methods.  
We note that for two RRab stars (those with the lowest $T_{\rm eff}$, 2156210476087226752 and 4446209728904885248) 
the best model from the RVS pipeline does not provide a good fit to the most prominent lines shown in Figure~\ref{fig:spectraMRYR}, and the measured [Fe/H] is mostly influenced by the presence of metallic lines in the reddest part of the spectra (namely, the calcium triplet region). 
Moreover, for these two stars the [Fe/H] measured by the SP pipeline and the $\Delta$S method are largely inconsistent with those from RVS (as shown in Table~\ref{tab:params_MRYR}).
Thus, we flag the high [Fe/H] status of these stars as tentative. 

As discussed by \citet{Bobrick2024}, although metallicity is the main discriminator between single and binary evolution-made RRLs, the detection of their binary companions is a confirmation of the latter. 
Hints of binarity in RRLs (i.e. evidence of an unseen companion) are typically inferred from the study of the observed minus calculated time of maximum brightness (O-C diagram), which is computed adopting a fixed period and known light curve ephemeris (in this case, a known time of maximum brightness).
Indeed, the change in distance between the observer and the RRL in a binary system over its motion around the center of mass of the system \citep[the light-travel time effect][]{Irwin1952,Irwin1959} induces a modulation of the light curves and therefore a deviation from the expected time of maximum brightness \citep[][]{Sterken2005}.
Thus, conducting studies of the O-C diagram of RRLs requires light curves spanning several years to measure cyclic period variations \citep[see e.g.,][]{Liska2016,Prudil2019,Hajdu2021,Abdollahi2025}. 
We note in passing that for this analysis we do not make special considerations for RRLs potentially exhibiting signatures of Blazhko effect. 
We advocate for a high-resolution spectroscopic follow up to confirm the metal-rich status of these stars, and a detailed analysis of their long-term variability to study indications of binary companions (which is beyond the scope of this work).

\begin{table*}\scriptsize
\caption{
Main properties of the metal-rich candidate RRLs ([Fe/H]$_{\rm RVS}>-0.5$\,dex) in our sample, including their {\it Gaia} {\tt source\_id}s, iron abundances, mean signal-to-noise ratio (S/N), heliocentric distance, and proper motions ($\mu_{\alpha^*}$, $\mu_\delta$). 
}
\label{tab:params_MRYR}
\begin{center}
\begin{tabular}{|c|c|c|c|c|c|c|c|c|c|c|}
\hline
\multicolumn{1}{|c|}{ID} &
\multicolumn{1}{c|}{Type} &
\multicolumn{1}{c|}{Period} &
\multicolumn{1}{c|}{${\rm Amp}_V$} &
\multicolumn{1}{c|}{[Fe/H]$_{\rm RVS}$} &
\multicolumn{1}{c|}{[Fe/H]$_{\rm SP}$} &
\multicolumn{1}{c|}{[Fe/H]$_{\Delta \rm S}$} &
\multicolumn{1}{c|}{S/N} &
\multicolumn{1}{c|}{$d_{\rm H}$} &
\multicolumn{1}{c|}{$\mu_{\alpha^*}$} &
\multicolumn{1}{c|}{$\mu_\delta$} \\     
 &  & (d) & (mag) & (dex) &   (dex) &  (dex) &  &  (kpc) & (mas\,yr$^{-1}$) &  (mas\,yr$^{-1}$)     \\     

\hline
2156210476087226752 & RRab & $0.6629$ & $0.40$ &   $-0.35 \pm 0.04$ &           $-1.63$ &                  -- &  $49$ & $11.4 \pm 0.2$ &  $-2.34 \pm 0.05$ &  $-4.81 \pm 0.05$ \\
4446209728904885248 & RRab & $0.5509$ & $0.39$ &   $-0.46 \pm 0.02$ &           $-1.76$ &    $-2.51 \pm 0.14$ &  $78$ &  $8.6 \pm 0.1$ &  $-5.36 \pm 0.04$ &   $1.71 \pm 0.03$ \\
4529262817181909376 & RRab & $0.4322$ & $1.05$ &   $-0.42 \pm 0.02$ &           $-0.45$ &    $-0.22 \pm 0.07$ &  $62$ &  $6.5 \pm 0.1$ &  $-2.55 \pm 0.02$ &  $-4.15 \pm 0.02$ \\
1865379961102699008 &  RRc & $0.3324$ & $0.31$ &   $-0.10 \pm 0.01$ &           $-0.63$ &    $-0.22 \pm 0.14$ & $109$ &  $2.7 \pm 0.0$ &   $0.55 \pm 0.01$ &  $-6.34 \pm 0.01$ \\
1316840138346618880 & RRab & $0.9292$ & $0.89$ &   $-0.48 \pm 0.01$ &           $-0.54$ &                  -- & $157$ &  $2.2 \pm 0.0$ &  $-6.26 \pm 0.01$ &  $-3.83 \pm 0.01$ \\
5734183009797858176 &  RRc & $0.3316$ & $0.13$ &   $-0.49 \pm 0.02$ &           $-0.83$ &    $-0.61 \pm 0.09$ &  $71$ &  $3.5 \pm 0.1$ &   $1.46 \pm 0.01$ &  $-2.96 \pm 0.01$ \\
1743670616671799296 & RRab & $0.4298$ & $1.11$ &   $-0.41 \pm 0.07$ &           $-0.70$ &    $-1.37 \pm 0.06$ &  $77$ &  $5.6 \pm 0.1$ &  $-5.51 \pm 0.03$ &  $-6.47 \pm 0.02$ \\
4556985044367410304 & RRab & $0.3905$ & $1.39$ &   $-0.26 \pm 0.02$ &           $-0.09$ &    $-1.26 \pm 0.21$ & $113$ &  $6.7 \pm 0.1$ &  $-0.13 \pm 0.02$ & $-10.01 \pm 0.03$ \\
\hline
\end{tabular}
\end{center}
\end{table*}

The orbital parameters of our metal-rich candidates, derived from our {\sc GALPY} integrations, are shown in Table~\ref{tab:orbital_params_MRYR}. 
Specifically, the table shows their galactocentric distances ($R_{\rm GC}$), total galactocentric velocities ($V$), energies ($E$), pericentric and apocentric distances of their orbits ($r_{\rm peri}$ and $r_{\rm apo}$, respectively), and the eccentricity of their orbits ($e$).

\begin{table*}\small
\caption{
Orbital parameters of the metal-rich RRL candidates in our sample, integrated for 5\,Gyr using GALPY's \textit{MWPotential2014} and taking into account the 
perturbation caused by the infall of the LMC. 
}
\label{tab:orbital_params_MRYR}
\begin{center}
\begin{tabular}{|c|c|c|c|c|c|c|cH|}
\hline
\multicolumn{1}{|c|}{ID} &
\multicolumn{1}{c|}{$R$} &
\multicolumn{1}{c|}{$V_{\rm tot}$} &
\multicolumn{1}{c|}{$E$} &
\multicolumn{1}{c|}{$L$} &
\multicolumn{1}{c|}{$r_{\rm peri}$} &
\multicolumn{1}{c|}{$r_{\rm apo}$} &
\multicolumn{1}{c}{$e$} &
\multicolumn{1}{H|}{Bound Likelihood}\\
  & (kpc) & (km\,s$^{-1}$)  &  ($10^{5}$\,km$^{2}$\,s$^{-2}$) &  ($10^{3}$\,kpc\,km\,s$^{-1}$) & (kpc) & (kpc) & &    \\    
\hline
2156210476087226752 & $14.0 \pm 0.2$ & $390.4^{+20.5}_{-18.3}$ & $-0.16^{+0.08}_{-0.07}$ & $5.38^{+0.26}_{-0.25}$ & $13.7^{+0.2}_{-3.4}$ & $70.4^{+9.7}_{-5.9}$ & $0.68^{+0.09}_{-0.04}$ &          $1.00$ \\
4446209728904885248 &  $6.0 \pm 0.1$ & $269.3^{+17.5}_{-12.8}$ & $-0.97^{+0.05}_{-0.04}$ & $0.18^{+0.11}_{-0.05}$ &  $0.3^{+0.2}_{-0.1}$ & $15.7^{+3.2}_{-2.2}$ & $0.97^{+0.01}_{-0.03}$ &          $1.00$ \\
4529262817181909376 &  $6.7 \pm 0.1$ & $132.4^{+18.4}_{-18.2}$ & $-1.25^{+0.03}_{-0.02}$ & $0.85^{+0.12}_{-0.12}$ &  $2.0^{+0.4}_{-0.4}$ &  $7.4^{+0.8}_{-0.5}$ & $0.58^{+0.09}_{-0.07}$ &          $1.00$ \\
1865379961102699008 &  $8.1 \pm 0.0$ &   $191.3^{+6.4}_{-6.9}$ & $-1.08^{+0.01}_{-0.01}$ & $1.51^{+0.05}_{-0.06}$ &  $3.6^{+0.8}_{-2.1}$ &  $9.9^{+1.1}_{-1.0}$ & $0.46^{+0.31}_{-0.09}$ &          $1.00$ \\
1316840138346618880 &  $7.5 \pm 0.0$ &   $143.1^{+4.2}_{-3.8}$ & $-1.19^{+0.01}_{-0.01}$ & $1.02^{+0.03}_{-0.03}$ &  $2.5^{+0.4}_{-0.4}$ &  $8.2^{+0.2}_{-0.3}$ & $0.54^{+0.06}_{-0.06}$ &          $1.00$ \\
5734183009797858176 & $10.4 \pm 0.1$ &   $215.9^{+3.8}_{-3.6}$ & $-0.89^{+0.01}_{-0.01}$ & $1.84^{+0.03}_{-0.02}$ &  $1.4^{+0.5}_{-0.8}$ & $19.5^{+2.2}_{-3.5}$ & $0.86^{+0.07}_{-0.06}$ &          $1.00$ \\
1743670616671799296 &  $7.7 \pm 0.1$ &   $170.3^{+3.3}_{-3.3}$ & $-1.11^{+0.01}_{-0.01}$ & $0.65^{+0.02}_{-0.01}$ &  $0.6^{+0.5}_{-0.3}$ & $11.1^{+0.6}_{-0.9}$ & $0.90^{+0.05}_{-0.08}$ &          $1.00$ \\
4556985044367410304 &  $6.7 \pm 0.1$ & $246.9^{+19.1}_{-12.0}$ & $-1.01^{+0.05}_{-0.03}$ & $0.72^{+0.11}_{-0.03}$ &  $0.5^{+0.4}_{-0.2}$ & $14.1^{+3.6}_{-2.2}$ & $0.94^{+0.03}_{-0.06}$ &          $1.00$ \\

\hline
\end{tabular}
\end{center}
\end{table*}

Based on the observed orbital properties of our sample, we identify four of the RRLs with ${\rm [Fe/H]} >-0.5$\,dex exhibiting disk-like kinematics (two thin disk and two thick disk), and four with halo-like orbits. 
The integrated orbits of the stars show a variety of shapes, with patterns that vary from case to case, and traits that include symmetric and asymmetric orbits, short and long orbital periods, among other properties. 
Figure~\ref{fig:orbitsMRYR} shows the orbital histories of four of these stars, including 10\% of the orbits generated to compute the orbital parameters and their uncertainties.
This figure shows the orbits of two stars that are mostly constrained to the disk region (at $Z<2$\,kpc) and two that reach larger apocentric distances. 
Among the latter, 4556985044367410304 displays a highly eccentric orbit that reaches $Z\sim10$\,kpc, with a pericenter of $0.5^{+0.4}_{-0.2}$\,kpc and an apocenter of $13.8^{+3.2}_{-1.9}$\,kpc. 
For the second halo-like star, 2156210476087226752, the integrated orbit suggests that it recently reached its pericenter (with $r_{\rm peri} = 16.7^{+0.5}_{-0.4}$\,kpc), and an apocenter of $\sim50$\,kpc. 
We note, however, that this RRL corresponds to one of the stars with a poor RVS model fit in the blue region of its spectrum (as discussed earlier in this section). 
If confirmed, the high metallicity of these stars and they observed parameters might provide hints of their origin, e.g., as indications of a dynamical ejection from the disk or through constraints on the nature of their formation (and potential association with binary companions).

In Figure~\ref{fig:LzvsE_HASP_SASP_MRYR}, the energy and vertical angular momentum of our sample is displayed, highlighting the position of the metal-rich candidates. 
The figure shows that these stars exhibit orbits that are predominantly prograde, consistent with their classification as disk-like stars. 
Indeed, the metal-rich candidates appear kinematically distinct to the bulk of the population (displaying radial orbits), in a region dominated by stars associated with the GSE merger event.
The position of the RRL with discrepant [Fe/H] measurements from different methods and 
large apocenter described above (2156210476087226752) in this diagram appears as an outlier of the overall distribution of metal-rich candidates, slightly below the $E\sim0$\,km$^{2}$\,s$^{-2}$ limit.

\section{Summary and conclusions}
\label{sec:conclusions}

The first year of operation of the Dark Energy Spectroscopic Instrument (DESI) Milky Way Survey provides one of the largest homogeneous samples of RR Lyrae stars (RRLs) available to date to investigate, empirically, existing correlations between their metallicities and their general physical properties. 
To this end, we used the 6,240 RRLs observed by DESI, with over 12,000 single-epoch observations and spectroscopic properties ([Fe/H], \teff) reported by \citet{Medina2025}, combined with the pulsating properties reported in the {\it Gaia} DR3 RRL catalog.  
The DESI RRL catalog is composed of field stars, and stars associated with known globular clusters, dwarf galaxies, and accretion events (most notably, Sagittarius and Gaia-Sausage-Enceladus).

We analyzed the metallicity gradients observed in the period-visual amplitude ($P-{\rm Amp}_V$) space (Bailey diagram).
For both fundamental and first-overtone mode RRLs, we find evidence of a strong and smooth anticorrelation between logarithmic period and iron abundances. 
In the case of RRc variables, these anticorrelations are more subtle if their visual amplitudes are considered instead of their periods, and for RRab stars the correlation is much less significant. 
With this information, we draw conclusions on the Oosterhoff dichotomy, a long-standing problem in modern astrophysics related to the metallicity of Galactic globular clusters and the mean period of the RRLs they contain, with potential implications on our understanding of the formation and evolution of the Milky Way, stellar evolution, and pulsation theory. 
Our measurements show smooth gradients in [Fe/H] with period, which support the findings of previous works regarding the Oosterhoff dichotomy as being caused by the dearth of 
intermediate metallicity 
Galactic globular clusters with sizeable RRL samples \citep[e.g.,][]{Fabrizio2019,Fabrizio2021}. 
Larger samples of DESI RRLs in globular clusters, e.g., taking advantage of available updated variable star membership to globular clusters \citep[][]{Prudil2024c} and large scale surveys will be important to provide additional evidence for or against this hypothesis, and the connection of the dichotomy with the dual origin of the Galactic halo \citep[see e.g.][]{Luongo2024}.

We identify stars in our catalog that fall into the high-amplitude short-period (HASP) and the small-amplitude short-period (SASP) categories (for RRab and RRc, respectively). 
Only field stars or stars associated with the Sagittarius stream populate the HASP and SASP regions in the Bailey diagram, and they are not observed in the sample of RRLs in the Draco dwarf galaxy, nor in the rest of dwarf galaxies and globular clusters observed by DESI Y1. 
This finding, combined with [Fe/H] distributions of HASP and SASP stars (with medians at $-1.39$\,dex and $-1.30$\,dex), is consistent with these stars having formed in massive satellites (with masses $>10^9$\,$M_\odot$; see e.g., \citealt{Fiorentino2017}) that experienced a fast early chemical enrichment and were subsequently accreted onto the Milky Way halo. 
We find a sizable subsample of HASPs and SASPs at large distances ($R_{\rm GC}>30$\,kpc), and advocate for a high-resolution spectroscopic follow up to gain further insight on their chemical abundances and their origin.

We studied the position of the $\sim100$ double mode (RRd) field stars in DESI Y1 in the period ratio vs. fundamental mode period space (Petersen diagram).  
In this space, which is a powerful diagnostic of the physical properties of old stellar population in stellar systems, we observe a smooth decline in [Fe/H] with increasing fundamental-mode period. 
Our sample includes several anomalous RRd stars, which exhibit a narrow range of [Fe/H] (between $-1.8$ and $-1.4$\,dex), in comparison with the [Fe/H] covered by our sample of classical RRd stars (between $-3.0$ and $-1.5$\,dex).
For the latter, we find that the first-overtone to fundamental period ratio is consistent with a linear [Fe/H] dependency. 
Our derived linear relation is mostly constrained to [Fe/H] between $-2.7$ and $-1.7$\,dex, and larger RRd samples are required to tests its validity outside this metallicity range.  
Lastly, using a theoretical mass-metallicity-period-ratio relation from the literature \citep{Marconi2015}, we constrain the mass of our classical RRd sample to $>0.69$\,M$_\odot$, and 0.68-0.77\,M$_\odot$ for anomalous RRd stars. 
These measurements highlight the importance of large spectroscopic surveys targeting sizeable samples of RRd stars (and their synergies with theoretical models) on the study of the pulsation properties of RRLs and stellar evolution.

The dependence of the instability strip topology on [Fe/H] provides important insights on the physical conditions required to trigger and maintain stellar pulsations, which influences the use of RRLs as stellar population tracers and the connection of their observed properties (e.g., RRc to RRab ratios in a given system, periods and period changes) with stellar pulsation theory.
Based on M25's \teff\ models, we set for the first time empirical constraints on the width of the RRL instability strip using a large spectroscopic dataset and pulsation-corrected \teff. 
For our halo field sample, we find that the red and blue edges of the instability strip lie at around  5963 and 7371\,K, respectively (with variations depending on metallicity).
We combined RRLs' mean \teff\ with DESI's RVS iron abundances to study the metallicity dependence of the instability strip's shape. 
We find a roughly constant \teff\ width of 1300-1400\,K as a function of [Fe/H]  and
an overall good agreement between our empirical estimates and those predicted from stellar evolution models \citep[from, e.g.,][]{Marconi2015}. 
However, we observe an instability strip that moves towards cooler \teff\ with declining [Fe/H] (similar to \citealt{Sandage1993} and \citealt{Luo2021}) in particular for its blue end, a trend that is not present in the theoretical estimates from \citet{Marconi2015}. 
This discrepancy could be explained by low number statistics and/or a decrease in precision in the derived [Fe/H] at the very metal-poor end, or the unsuitability of the theoretical estimates at low metallicities. 
We also inspect the transition \teff\ region between RRab and RRc stars, where most of the RRd are found, which occurs at $\sim6440$-$6700$\,K (depending on metallicity).

Taking advantage of the large DESI catalog, we inspected the metal-rich end of the RRL [Fe/H] distribution, given the implications of metal-rich RRL formation on stellar evolution theory, binarity, and the use of RRLs as tracers of old stellar populations only.  
We find eight RRLs 
with ${\rm [Fe/H]}>-0.5$\,dex, as measured by DESI's main processing pipeline (RVS). 
Of these stars, two exhibit similarly high metallicities 
derived by DESI's SP pipeline or the $\Delta$S method, and one shows a high metallicity from all three methods. 
In order to gain insight on the dynamical origin of this subsample, we integrate their orbits using their systemic (center-of-mass) velocities combined with their {\it Gaia} proper motions. 
Our results suggest disk-like kinematics for half of the stars, whereas the other half displays halo kinematics.
Further investigations are required to draw firmer conclusions on the origin of these stars through indications of binarity (one of the configurations that can give rise to metal-rich RRLs; see e.g., \citealt{Bobrick2024}), and as for the HASP and SASP samples, we advocate for a high-resolution spectroscopic follow up and/or multi-epoch spectra to confirm their observed high metallicities and orbital histories.

The spectroscopic characterization of RRLs  to study their formation conditions and their connection with the assembly history of the halo requires large databases containing homogeneously-derived spectroscopic properties, which must take into account their short-period cyclic variations. 
The next generation of facilities and instruments is delivering (and will deliver) such databases \citep[e.g., the Subaru Prime Focus Spectrograph survey and the 4-metre Multi-Object Spectroscopic Telescope;][]{deJong2014,Takada2014,Tamura2016,Ibata2023_4grounds}, including RRLs in all regions of the Galaxy, in globular clusters, and dwarf galaxies. 
Furthermore, future DESI data releases will significantly increase the observed number of RRLs in the halo, the number of epochs per star, and will improve the variation models used to derive their main properties (e.g., their systemic velocities, effective temperatures, and [Fe/H]).
These large samples of RRLs will also enable the exploration of multiple science cases not covered in this work, including but not limited to, 
the study of metallicity gradients in globular clusters and dwarf galaxies and the identification of substructures in integrals-of-motion space.  
Thus, especially when combined with upcoming {\it Gaia} data releases and their improvements in astrometric precision and accuracy, these databases will give us the tools to investigate, empirically, multiple aspects of Galaxy evolution, stellar astrophysics, and pulsation theory with unseen detail.

\section*{Data availability}
The data shown in all the figures of this manuscript will be made available in Zenodo: \href{https://doi.org/10.5281/zenodo.15347442}{https://doi.org/10.5281/zenodo.15347442}.

\acknowledgments
We thank the anonymous referee for their detailed and thoughtful review, whose comments and suggestions substantially improved this work.
We acknowledge very useful feedback from
M\'arcio Catelan (Pontificia Universidad Cat\'olica de Chile), 
Marcella Marconi (Istituto Nazionale di Astrofisica, INAF, Osservatorio Astronomico di Capodimonte), and Vittorio Braga (Istituto Nazionale di Astrofisica, INAF, Osservatorio Astronomico di Roma).
G.E.M. and T.S.L. acknowledge financial support from Natural Sciences and Engineering Research Council of Canada (NSERC) through grant RGPIN-2022-04794.
G.E.M. acknowledges support from an Arts \& Science Postdoctoral Fellowship and Dunlap Fellowship at the University of Toronto. 
The Dunlap Institute is funded through an endowment established by the David Dunlap family and the University of Toronto.
S.K. acknowledges support from Science \& Technology Facilities Council (STFC) (grant ST/Y001001/1).

This material is based upon work supported by the U.S. Department of Energy (DOE), Office of Science, Office of High-Energy Physics, under Contract No. DE–AC02–05CH11231, and by the National Energy Research Scientific Computing Center, a DOE Office of Science User Facility under the same contract. Additional support for DESI was provided by the U.S. National Science Foundation (NSF), Division of Astronomical Sciences under Contract No. AST-0950945 to the NSF’s National Optical-Infrared Astronomy Research Laboratory; the Science and Technology Facilities Council of the United Kingdom; the Gordon and Betty Moore Foundation; the Heising-Simons Foundation; the French Alternative Energies and Atomic Energy Commission (CEA); the National Council of Humanities, Science and Technology of Mexico (CONAHCYT); the Ministry of Science, Innovation and Universities of Spain (MICIU/AEI/10.13039/501100011033), and by the DESI Member Institutions: \url{https://www.desi.lbl.gov/collaborating-institutions}. Any opinions, findings, and conclusions or recommendations expressed in this material are those of the author(s) and do not necessarily reflect the views of the U. S. National Science Foundation, the U. S. Department of Energy, or any of the listed funding agencies.

The authors are honored to be permitted to conduct scientific research on Iolkam Du’ag (Kitt Peak), a mountain with particular significance to the Tohono O’odham Nation.

% openaccess requirement 
For the purpose of open access, the author has applied a Creative Commons Attribution (CC BY) licence to any Author Accepted Manuscript version arising from this submission.

% SIMBAD
This research has made use of the SIMBAD database, operated at CDS, Strasbourg, France \citep{Simbad}.
This research has made use of NASA’s Astrophysics Data System Bibliographic Services.

{\it Software:} 
{\code{numpy} \citep{numpy}, 
\code{scipy} \citep{2020SciPy-NMeth},
\code{matplotlib} \citep{matplotlib}, 
\code{astropy} \citep{astropy,astropy:2018, astropy:2022},
\code{galpy} \citep{Bovy2015}, 
\code{RVSpecFit} \citep{rvspecfit},
 \code{emcee} \citep{emcee}.
}

\appendix
\renewcommand\thefigure{\thesection\arabic{figure}}    
\setcounter{figure}{0}    

%\counterwithin{table}{section}
\renewcommand{\thetable}{\thesection\arabic{table}}
\setcounter{table}{0}

\section{Trapezoid model: posterior distributions and inferred intrinsic profiles}
\label{sec:appendixA1}

In this section, we provide the corner plots corresponding to the parametric representation of the instability strip of RRab and RRc stars described in Section~\ref{sec:trapezoid}, and the trapezoidal probability density $p_{\mathrm{int}}(T_{\rm eff} \mid {\rm [Fe/H]},\,\boldsymbol{\theta})$
(Equation~\ref{eq:IS_trapezoid4}) at ${\rm [Fe/H]}=-1.5$ for RRab and RRc stars.

\begin{figure*}
\begin{center}
\includegraphics[angle=0,scale=.25]{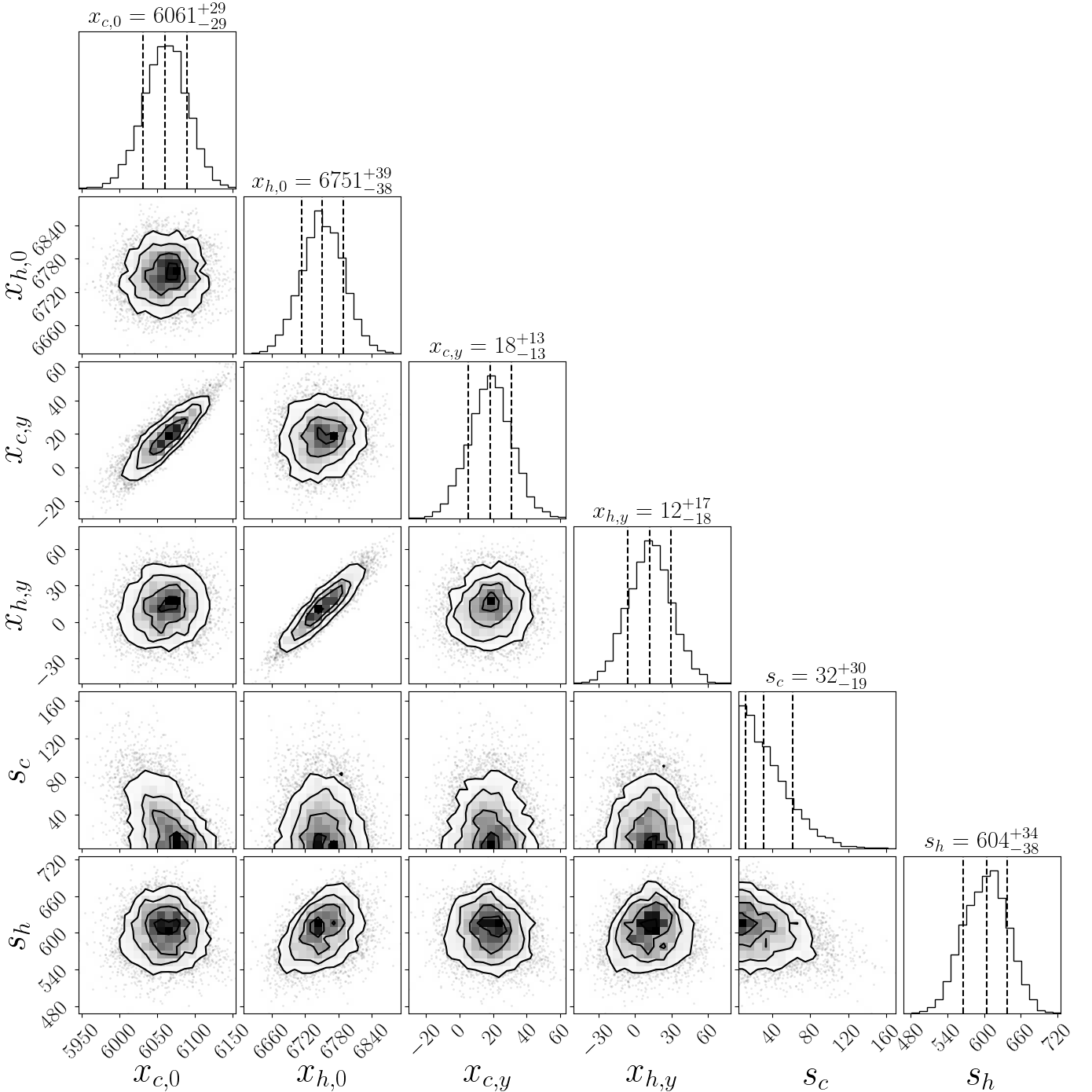}
\includegraphics[angle=0,scale=.25]{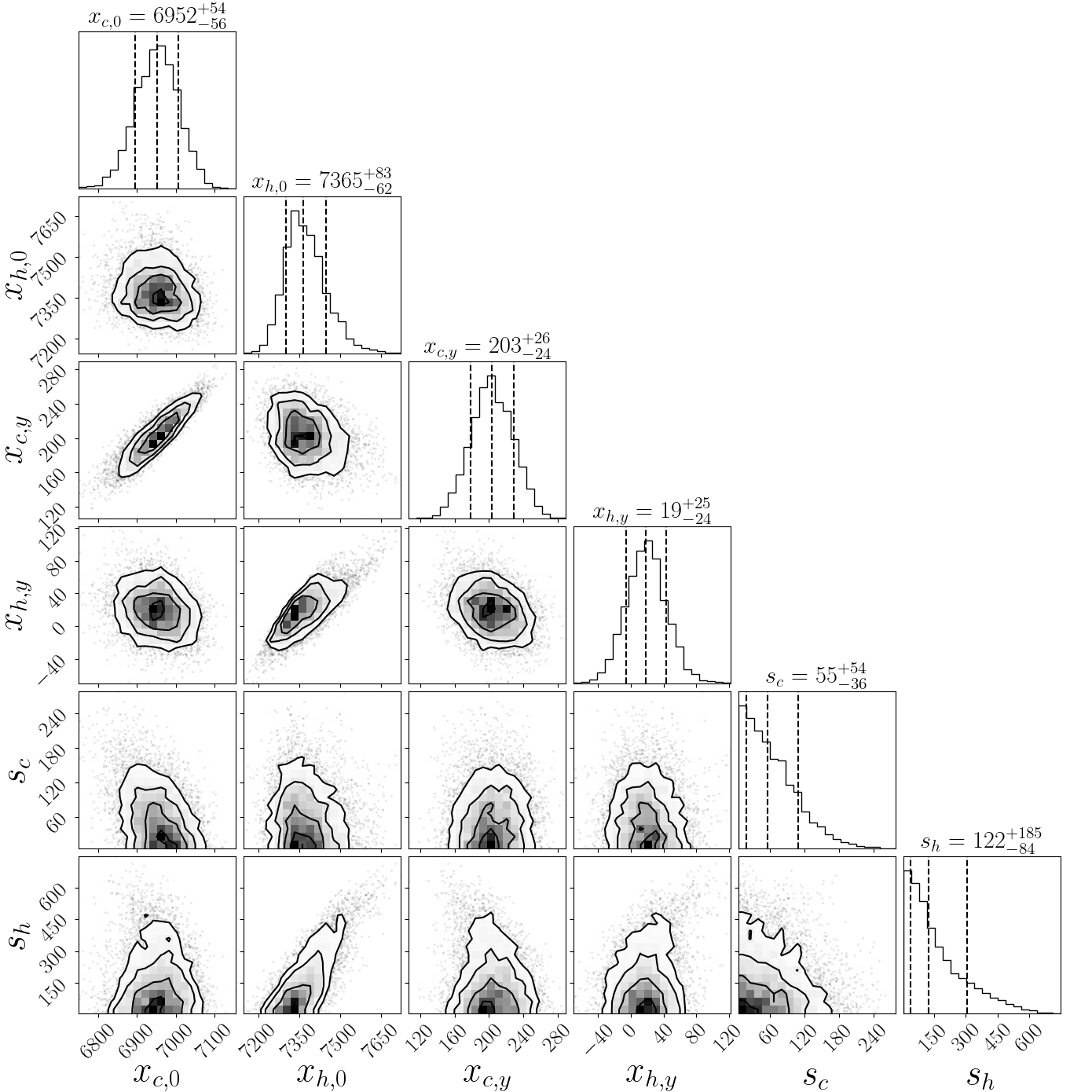}
\caption{
Corner plots of the parameters used to model the \teff\ distribution of RRab stars (left) and RRc stars (right).
}
\label{fig:cornerplots_trapezoid}
\end{center}
\end{figure*}

\begin{figure}
\begin{center} 
\includegraphics[angle=0,scale=.43]{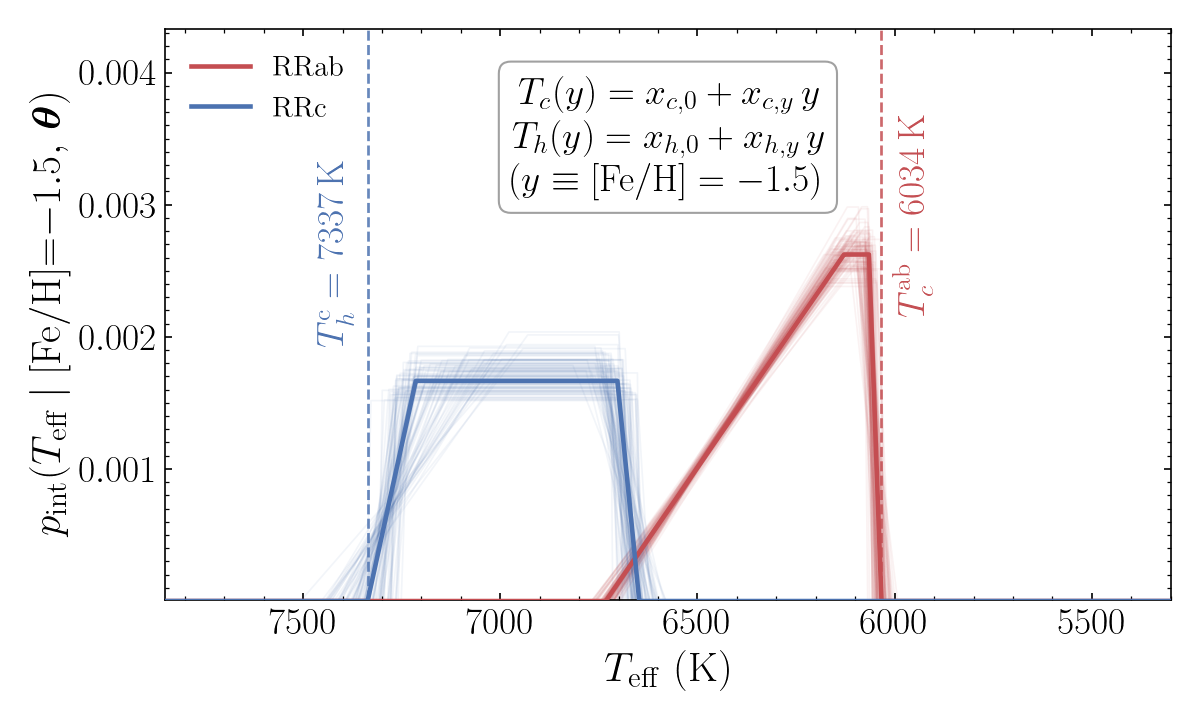}
\caption{
Intrinsic trapezoidal probability density 
$p_{\mathrm{int}}(T_{\rm eff} \mid {\rm [Fe/H]},\,\boldsymbol{\theta})$
(Equation~\ref{eq:IS_trapezoid4}) at ${\rm [Fe/H]}=-1.5$ for RRab (red) and RRc (blue) stars.
The solid curves show the median-parameter model.
The 100 transparent curves depict random draws from the MCMC posterior chains, illustrating the uncertainty in the inferred intrinsic distributions.
}
\label{fig:trapezoid_realizations}
\end{center} 
\end{figure}

\newpage

\bibliography{references} % if your bibtex file is called example.bib
\bibliographystyle{aasjournal}


% End of file `sample631.tex'.

\end{document}